# Understanding and Minimizing $V_{OC}$ Losses in All-Perovskite Tandem Photovoltaics

*Jarla Thiesbrummel[1,2,\*], Francisco Peña-Camargo[1], Kai Oliver Brinkmann[3], Emilio Gutierrez-Partida[1], Fengjiu Yang[4], Jonathan Warby[1], Steve Albrecht[4,5], Dieter Neher[1], Thomas Riedl[3], Henry J. Snaith[2,\*], Martin Stolterfoht[1,\*], Felix Lang[1,\*]*

[1] Institute of Physics and Astronomy University of Potsdam Karl-Liebknecht-Str. 24–25, 14476 Potsdam-Golm, Germany

[2] Clarendon Laboratory, University of Oxford, Parks Road, Oxford, OX1 3PU UK

[3] Institute of Electronic Devices and Wuppertal Center for Smart Materials & Systems, University of Wuppertal, Rainer-Gruenter-Str. 21, 42119 Wuppertal, Germany

[4] Young Investigator Group Perovskite Tandem Solar Cells, Helmholtz-Zentrum Berlin für Materialien und Energie GmbH, Kekuléstraße 5, 12489 Berlin, Germany

[5] Technical University Berlin, Faculty IV – Electrical Engineering and Computer Science, Marchstr. 23, 10587 Berlin, Germany

\* Emails

jarla.thiesbrummel@physics.ox.ac.uk, henry.snaith@physics.ox.ac.uk, stolterf@uni-potsdam.de, felix.lang.1@uni-potsdam.de

**Broader context**
Combining two different perovskite thin-films, which absorb light over different regions of the solar spectrum, into tandem solar cells, presents an elegant way to improve the efficiency, energy yield, and energy-payback time, rendering them highly attractive solutions to accelerate the transition to renewable energy systems. Optimization of such interwoven systems however is complex and, once assembled into a "monolithic" tandem stack, the performance of the individual subcells cannot be assessed directly. This results in missing feedback parameters, a key problem slowing down progress made. In this work, we demonstrate a thorough subcell diagnosis methodology that provides deep insights into the performance and limiting loss mechanisms of the individual perovskite subcells when assembled in a tandem solar cell. We identify and eliminate dominating voltage losses in the high bandgap subcell and then present a thorough analysis on optimized systems identifying further routes for optimization.

## Abstract
All-perovskite tandem solar cells promise high photovoltaic performance at low cost. So far however, their efficiencies cannot compete with traditional inorganic multi-junction solar cells and they generally underperform in comparison to what is expected from the isolated single junction devices. Understanding performance losses in all-perovskite tandem solar cells is a crucial aspect that will accelerate advancement. Here, we perform extensive selective characterization of the individual sub-cells to disentangle the different losses and limiting factors in these tandem devices. We find that non-radiative losses in the high-gap subcell dominate the overall recombination losses in our baseline system as well as in the majority of literature reports. We consecutively improve the high-gap perovskite subcell through a multi-faceted approach, allowing us to enhance the open-circuit voltage ($V_{OC}$) of the subcell by up to 120 mV. Due to the (quasi) lossless indium oxide interconnect which we employ for the first time in all-perovskite tandems, the $V_{OC}$ improvements achieved in the high-gap perovskites translate directly to improved all-perovskite tandem solar cells with a champion $V_{OC}$ of 2.00 V and a stabilized efficiency of 23.7%. The efficiency potential of our optimized all-perovskite tandems reaches 25.2% and 27.0% when determined from electro- and photo-luminescence respectively, indicating significant transport losses as well as imperfect energy-alignment between the perovskite and the transport layers in the experimental devices. Further improvements to 28.4% are possible considering the bulk quality of both absorbers measured using photo-luminescence on isolated perovskite layers. Our insights therefore not only show an optimization example but a generalizable evidence-based strategy for optimization utilizing optical sub-cell characterization.

## Introduction

With the discovery that mixed-metal halide perovskites enable much lower bandgaps than their neat-lead or neat-tin based counterparts, significant efforts commenced on the development of all-perovskite tandem solar cells [1–5]. Combining low-band gap and high-band gap perovskites in tandem solar cells overcomes the fundamental efficiency limits of their single junction counterparts, without the need for combining with more traditional low-bandgap materials used previously such as Si or copper indium gallium selenide (CIGS) [6–10]. All-perovskite tandems promise highest efficiencies, on par with perovskite/silicon tandem technologies, while using much thinner absorber layers, and move away from the energy intensive production of crystalline silicon. They can also be much lighter, which makes them a promising option for a range of different applications: from building- or vehicle-integrated PV to high-altitude and even space PV where they also benefit from their radiation hardness [11]. Their energy-efficient processability, either from solution or by thermal evaporation at ambient temperatures, is roll-to-roll compatible and could allow a much more cost-efficient technology with shorter energy payback times compared to current technologies, or even perovskite/Si tandem PV [12].

The solar to electrical power conversion efficiency for 2-Terminal all-perovskite tandems has increased from 17% for the first attempts, to the current record of 26.4% [3,13,14]. We give an overview of this rapid progress in **Figure 1a,** by plotting the employed bandgap combinations and the achieved power conversion efficiencies, alongside with a realistic efficiency potential of ~ 0.75 × PCE[rad]. Here PCE[rad] is the radiative efficiency from the detailed balance limit assuming a step function absorption profile for each band gap (see supplementary information for calculation details). A large focus has been on improving the efficiency and stability of the low bandgap lead-tin perovskite [13,15–18]. However, as we illustrate in **Figure 1b**, the high-gap perovskites dominate the open-circuit voltage ($V_{OC}$) losses for most tandem devices, especially after significant improvement of low bandgap perovskites over the past three years.

Overcoming the $V_{OC}$ losses in the high-gap perovskite subcell, which typically increase for higher bandgap perovskites [19–21], therefore offers room for further improvements. Moreover, incorporation of the high gap and low gap perovskites in a tandem often leads to significant additional $V_{OC}$ penalties. This can be due to degradation of the underlying subcell or layers during subsequent layer deposition, be it via sputtering or from solution, additional interfacial recombination induced by the interconnecting layer, or processing issues and inhomogeneities. We highlight several of those cases with a 'D' in **Figure 1b**.

To understand where these losses come from and how they can be reduced, it is important to not just look at the overall tandem performance and single junction cells, but characterize the behavior of both subcells when incorporated in the complete tandem device. Traditional electrical characterization of the monolithic tandem, however, provides little information on the behavior of the individual subcells. *JV*-measurements performed on corresponding single junction devices, that are often reported alongside tandem results and used in **Figure 1b,** do not necessarily reflect subcell performance once integrated in the tandem accurately. Quantitative measurements that provide information on the different subcells within the monolithic tandem stack individually are therefore crucial to understand performance-limiting layers and mechanisms. Electro- and photo-luminescence (EL & PL) can be measured from each sub-cell selectively in monolithic interconnected tandem devices, and we recently used this approach to reveal efficiency limits in perovskite/silicon tandems [22].

Herein, we conduct extensive PL and EL characterization of isolated perovskite films, single junction devices, and complete all-perovskite tandem cells, in order to identify the factors limiting the performance in these cells in comparison with the thermodynamic efficiency limit. We reveal that the high-band gap sub cell is predominantly responsible for the $V_{OC}$ losses in our own complete tandem devices, as well as in many tandem devices from literature at the moment. We employ a three-fold optimization strategy for the high-band gap perovskite, consisting of; i) addition of oleylamine to the perovskite in combination with, ii) a lithium fluoride (LiF) layer between the perovskite and the electron transport layer (ETL) and, iii) the use of the self-assembled monolayer (SAM) 2PACz instead of the frequently used hole transport layer (HTL) poly[bis(4-phenyl)(2,4,6-trimethylphenyl)amine] (PTAA). We apply our three-fold optimization approach to triple cation-based high-gap perovskites with bandgaps ranging from 1.80 through 1.85 to 1.88 eV and find a robust reduction of $V_{OC}$ losses for all tested bandgaps. This is important since the latter two bandgaps promise highest power conversion efficiencies in combination with the 1.27 eV perovskite used herein (**Figure 1b**). We then use the optimized high-gap perovskites to fabricate efficient all-perovskite tandem solar cells, reaching steady-state efficiencies of up to 23.4, 23.7 and 21.5% for 1.80 eV/1.27 eV, 1.85 eV/1.27 eV and 1.88 eV/1.27 eV bandgap combinations, respectively. Coming back to the subcell characterization, we then characterize these optimized all-perovskite tandems, and are able to determine the efficiency potential that could be achieved if transport losses and energy-level mismatches in the stack were eliminated. Furthermore, we also show that the interconnect we employed is lossless. Overall, our versatile sub-cell characterization approach will facilitate evidence-based optimization of future tandem cells.

## Results
**Assessing the limiting junction in the tandem cells**

In order to investigate whether the $V_{OC}$ losses in our all-perovskite tandems are dominated by the high- or the low-gap perovskite subcells, we measured the photoluminescence quantum yield (PLQY) selectively by excitation with 520 nm and 818 nm in a monolithic all perovskite tandem (based on a 1.27 eV low-gap and 1.80 eV high-gap combination with an efficiency of 20.9% under AM1.5G) and compared it to the PLQY of identically prepared isolated high-gap (PLQY = 0.1%) and low-gap (PLQY = 0.98%) perovskite layers on glass (**Figure 1c**). Interestingly, the PLQY of the high-gap perovskite is reduced by three orders of magnitude in the tandem device, compared to the isolated layer whereas the low-gap perovskite is only reduced ~20-fold. Since the quasi-Fermi level splitting (QFLS) is directly given by the PLQY and the radiative limit of the semiconducting material ($QFLS_{rad}$, see supplementary information) via equation (1), we conclude that the high-gap perovskite strongly dominates $V_{OC}$ losses also in our system.

$$\mathrm{QFLS} = \mathrm{QFLS_{rad}} + k_B T \cdot \ln(\mathrm{PLQY}) \quad (\text{eq. 1})$$

To further understand the potential of our tandem with the given absorbers, we now measure intensity-dependent photoluminescence yields (iPLQY) which allows us to determine a QFLS at each intensity. We then construct 'pseudo-*JV* curves' by plotting the total recombination current at each excitation intensity minus the generation current ($J_{SC}$) on the y-axis versus the QFLS on the x-axis (**Figure 1d**). Comparison of pseudo-*JV* curves derived from iPLQY measurements on isolated films vs. measurements in the monolithic tandem stack exemplifies $V_{OC}$ losses present in the tandem configuration. It also shows that these losses are dominated by the high-gap perovskite. Summing the QFLS obtained for the high-gap and low-gap perovskite isolated layers and subcells, further allows us to construct pseudo-*JV* curves of corresponding tandems that are free of resistive losses and, in case of the isolated layers, additionally free of interface recombination from the various contact layers, processing damage, etc. This efficiency potential constructed from isolated layer measurements reaches 28.2 %, a value which corresponds to the practical efficiency potential of around 0.75× $PCE^{rad}$. We present a more detailed analysis as well as strategies to reach this potential at the end of this work.

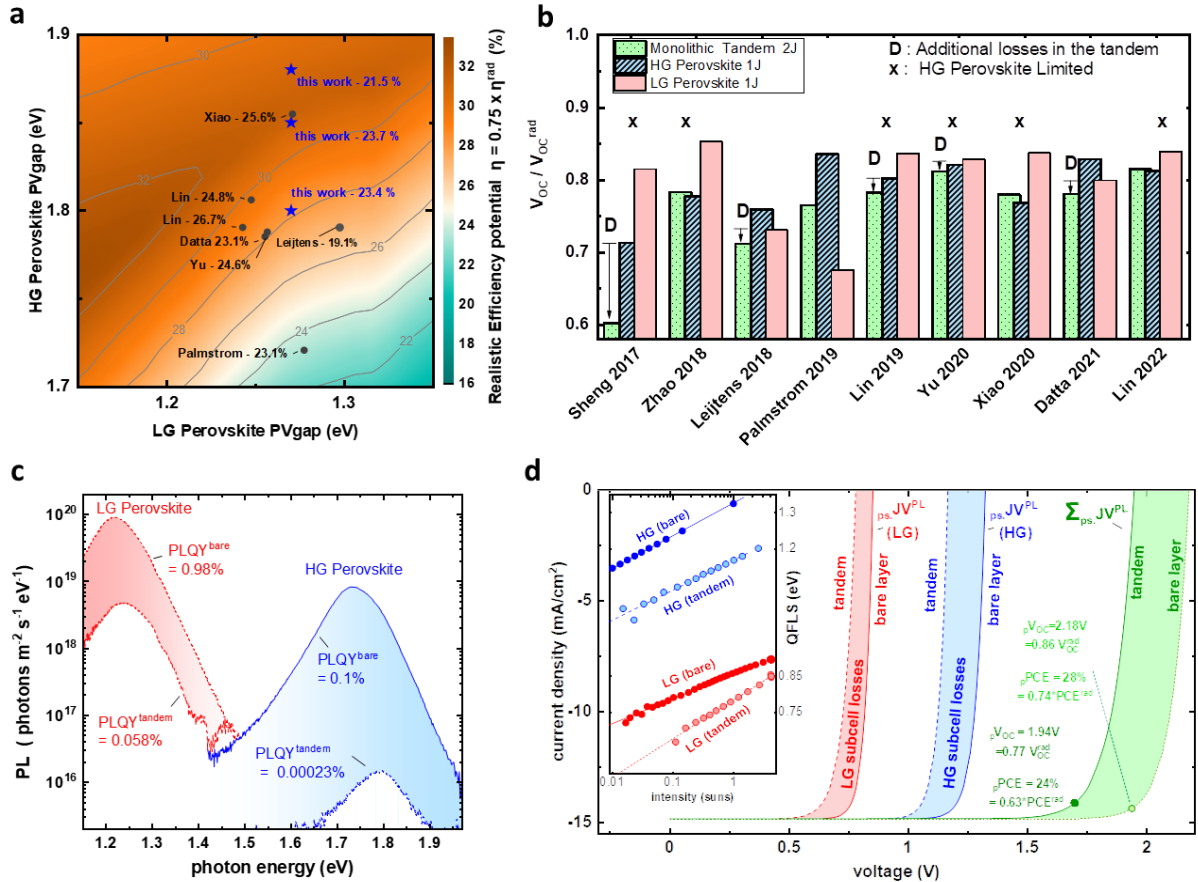

**Figure 1.** a) Literature overview of various all-perovskite tandem solar cells indicating the achieved efficiency, plotted as a function of the employed high-and low bandgap perovskites. For consistency, we determined the individual bandgaps from reported external quantum efficiencies (EQE), via d(EQE)/dE, and denoted this as PVgap. The color map displays the realistic efficiency potential for specific bandgap combinations (defined as 75% of the radiative efficiency limit). b) Relative $V_{OC}$ losses in monolithic all-perovskite tandems from a), as well as the corresponding high-gap and low-gap single junction devices [13,16,18,23–29]. c) Photoluminescence (PL) of high- and low-gap perovskites fabricated individually on glass (isolated bare layers) and incorporated in a monolithic tandem. d) Pseudo-*JV* curves reconstructed from intensity-dependent PL measurements highlighting $V_{OC}$ losses.

## Minimization of $V_{OC}$ losses in HG Perovskites

To understand the origins of $V_{OC}$ limitations in our high-gap perovskite we measured the PLQY of 1.80 eV bandgap triple-cation based perovskites ($Cs_{0.05}(FA_{0.60}MA_{0.40})_{0.95}Pb(I_{0.60}Br_{0.40})_3$) with and without the hole-and electron transport layers (HTL and ETL respectively). Comparing the PLQY of a perovskite layer prepared on glass and on PTAA, our standard HTL, as displayed in **Figure 2a**, reveals that the PTAA is strongly limiting the PLQY. This limitation is known and has been addressed in the past by using self-assembled monolayer (SAM) HTLs instead of conventional HTLs [30,31]. Their use as a hole selective contact can strongly reduce non-radiative losses at the perovskite – HTL interface compared to the conventionally used PTAA, [30] which is particularly a problem for wide-gap perovskites (>1.75 eV) [20].

Changing the HTL from PTAA to 2PACz significantly improves the PLQY of our 1.80 eV triple-cation based perovskites (**Figure 2a**), indicating a reduction of interfacial recombination at the HTL side, which was strongly limiting before. Having removed the dominating HTL-Pero-interface limitation now allows us to address the perovskite-ETL interface. We first try inserting a thin layer of LiF between the perovskite and the ETL, forming a strong surface dipole which repels minority carriers away from the interface, thereby reducing interfacial recombination[31–33]. In order to reduce the recombination losses further, we then added oleylamine into the perovskite precursor solution, which has previously been shown to improve the efficiency of lead-based perovskite solar cells through both grain- and interface modifications[34]. Interestingly when we tested a combined approach of adding oleylamine to the perovskite precursor and inserting a thin LiF layer we found that they worked in an additive fashion, increasing the QFLS to exactly the sum of the individual improvements (**Figure 2a)**.

We subsequently fabricated single junction solar cells with the structure glass/ITO/PTAA or 2PACz/1.80 eV bandgap triple-cation based perovskite w/ or w/o oleylamine /ETL w/ or w/o LiF /Cu to test our three-fold optimization approach prior to its incorporation into tandem devices. Looking at the *JV* curves displayed in **Figure 2b**, it can be seen that the improvement in PLQY directly translates into an increased $V_{OC}$ in devices. As seen in Table 1, summarizing device parameters and statistics (more statistics can be found in **Figure S1**), the $V_{OC}$ improves about 90 mV for the fully optimized device compared to control devices, which equals the sum of the individual gains. **Figure 2c** shows that every step of our three-fold optimization brings the $V_{OC}$ significantly closer towards the limit imposed by the bulk quality of the perovskite absorber, which is 1.32 V as determined from PLQY measurements of isolated perovskite absorbers. Notably, the three-fold optimized devices reach an average $V_{OC}$ of 1.26 V (max 1.29 V) which is around 83% of the radiative $V_{OC}$ limit of 1.51 V for devices with a 1.80 eV bandgap.

We tested the robustness of this three-fold optimization route with various high-bandgap perovskite compositions by varying the Br-ratio from the initial 0.4 to 0.45 and 0.5. This allowed us to vary the perovskite bandgap from 1.80 eV to 1.85 and 1.88 eV respectively, as determined from EQE. Control devices based on the 1.85 and 1.88 eV perovskites both reached 1.17 V $V_{OC}$ on average, see **Figure S2**. Using our three-fold optimization significantly improved the $V_{OC}$ to 1.27 V and 1.28 V, thereby achieving remarkable $V_{OC}/V_{OC}^{rad}$ ratios of 0.82 and 0.81, respectively, as can be seen in **Figure 2d**. This improvement becomes even more apparent when comparing our achieved $V_{OC}$ with data obtained from literature. In **Figure 2e** we display the ratio of the $V_{OC}$ divided by the $V_{OC}^{rad}$ as a function of the perovskite bandgap for a large number of perovskite *pin* devices, extracted from *The Perovskite Database* [21]. Clearly visible is a general trend of decreasing $V_{OC}/V_{OC}^{rad}$ with increasing bandgap [35]. This effect has been assigned to different phenomena, from halide segregation (or Hoke effect), to interface recombination, improper energy alignment, or high defect densities at the surface of the perovskite[19,20,36].

Our optimized 1.80, 1.85 and 1.88 eV perovskite based single junction devices (blue stars) thereby reach comparatively high $V_{OC}/V_{OC}^{rad}$ ratios, well above 0.8. This highlights that our three-fold optimization route, that was initially developed for the HG perovskite with a bandgap of 1.80 eV, is applicable to a wider range of perovskite compositions. A robust passivation strategy is critical for future all-perovskite tandem development to unlock highest efficiencies with optimal HG - LG bandgap combinations, as shown in **Figure 1a.**

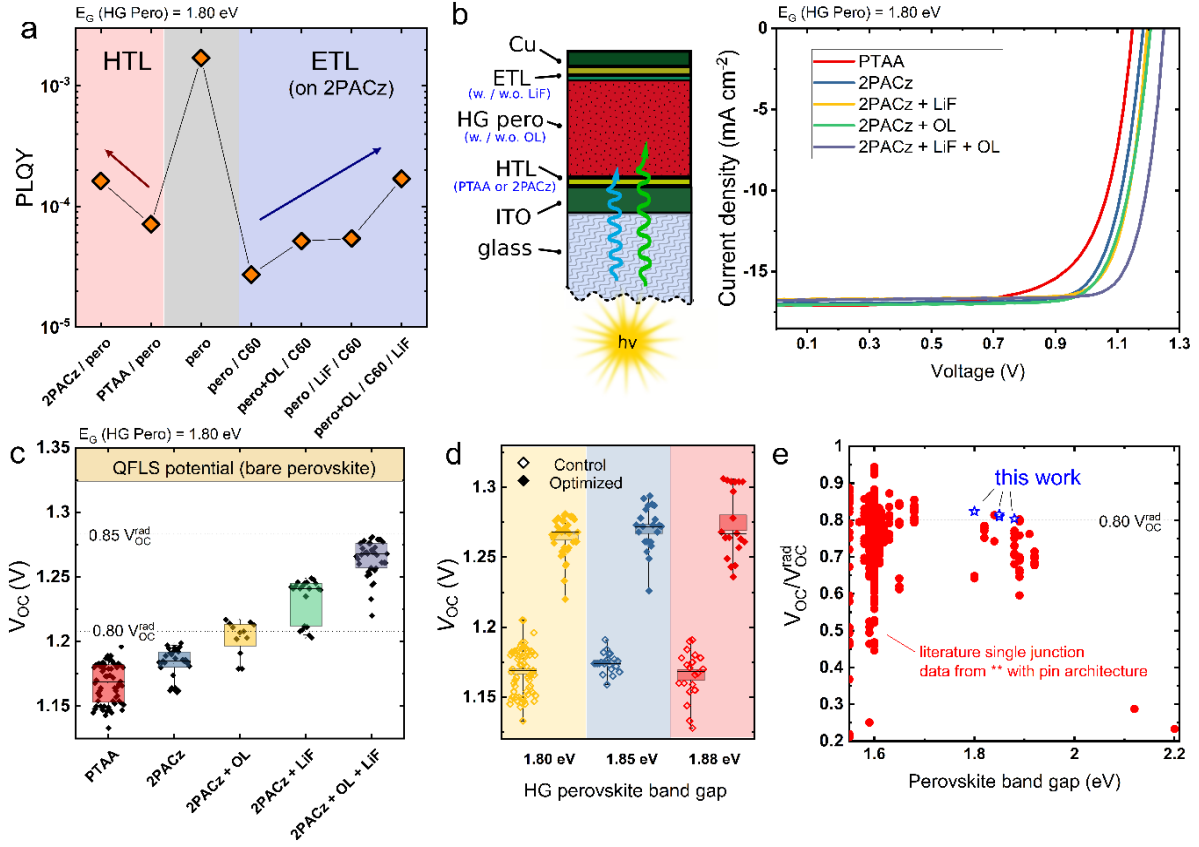

**Figure 2.** a) PLQY measured on different partial stacks. b) Schematic overview of the HG single junction solar cell and JV curves for 1.80 eV $Cs_{0.05}(FA_{0.60}MA_{0.40})_{0.95}Pb(I_{0.60}Br_{0.40})_3$ perovskite solar cells for various optimization steps. c) The $V_{OC}$ from the solar cells in (b), plotted as a function of the different optimization steps. The dashed line displays the radiative $V_{OC}$ limit ($V_{OC}^{rad}$), which is 1.51 V for perovskites with a 1.80 eV bandgap. d) $V_{OC}$ statistics for optimized and control HG perovskite solar cells with bandgaps of 1.80, 1.85 and 1.88 eV. e) The $V_{OC}$ divided by the radiative $V_{OC}$ limit for recent pin perovskites with a range of different bandgaps, extracted from The Perovskite Database [21]. The stars indicate our work and show optimized (champion) solar cells based on perovskites bandgaps at 1.80, 1.85, and 1.88 eV.

|  | $V_{OC}$ [V] | $J_{SC}$ [mA cm$^{-2}$] | FF [%] | PCE [%] | $\Delta V_{OC}$ [mV] | $\Delta FF$ [%] |
|---|---|---|---|---|---|---|
| PTAA device | 1.17 ± 0.02 | 17.0 ± 0.5 | 75.6 ± 2.9 | 15.0 ± 1.0 | - | - |
| 2PACz device | 1.18 ± 0.01 | 17.2 ± 0.7 | 78.8 ± 1.4 | 16.1 ± 0.6 | 10 | 3.2 |
| 2PACz + OAm | 1.20 ± 0.01 | 17.1 ± 0.2 | 77.2 ± 2.0 | 15.9 ± 0.7 | 30 | 1.6 |
| 2PACz + LiF | 1.23 ± 0.02 | 17.1 ± 0.5 | 78.9 ± 2.0 | 16.6 ± 0.6 | 60 | 3.3 |
| 2PACZ + OAm and LiF | 1.26 ± 0.01 | 16.6 ± 0.5 | 79.1 ± 2.0 | 16.6 ± 0.6 | 90 | 3.5 |

**Table 1.** Device parameters and statistics for the different optimized and control single junction devices using a 1.80 eV HG perovskite absorber. $\Delta V_{OC}$ And $\Delta FF$ indicate the improvement in these respective values compared to the unoptimized control device on PTAA.

## Implementation in Tandems

In a next step we integrated our three-fold optimized high gap perovskites into monolithic all perovskite tandems. For this purpose, we use a 1.27 eV (as determined by d(EQE)/dE) low-gap $FA_{0.83}Cs_{0.17}Pb_{0.5}Sn_{0.5}I_3$ perovskite subcell, which we deposited on top of the high-gap subcells. The complete layer stack comprises glass/ITO/PTAA or 2PACz/HG-Perovskite/$C_{60}$/ AZO-nanoparticles/ALD-SnOx/ALD-InOx/PEDOT:PSS/LG-Perovskite/$C_{60}$/BCP/copper as shown in **Figure 3a** alongside a cross-sectional scanning electron microscopy image of the all-perovskite tandem structure. Our recombination layer comprises an ultrathin (~ 1.5nm) layer of indium oxide, like we have reported previously for perovskite/organic tandem solar cells [37]. This indium oxide layer is deposited by atomic layer deposition (ALD) on top of a hybrid AZO-NP / ALD $SnO_x$ layer that functions as internal barrier layer, stabilizing and protecting the layers underneath from follow up processing [38,39]. This is the first time that such an interconnect is implemented in a perovskite/perovskite tandem structure.

As seen in the *JV* characteristics in **Figure 3b**, our all-perovskite tandems with optimized HG perovskite sub-cells reach a much higher $V_{OC}$, than the non-optimized control tandems. The improvement in $V_{OC}$, of about 120 mV, is consistent with the improvement observed in the optimized HG single junctions, and we present more detailed sub-cell analysis later on. Optimized tandems reach 78% of their radiative $V_{OC}$ limit, see **Figure 3c** and maximum power point (MPP) tracking, displayed in the inset of **Figure 3b**, shows that the optimized all-perovskite tandems reach PCE's of 23.4%, an improvement of about 2.5% absolute compared to the control tandems.

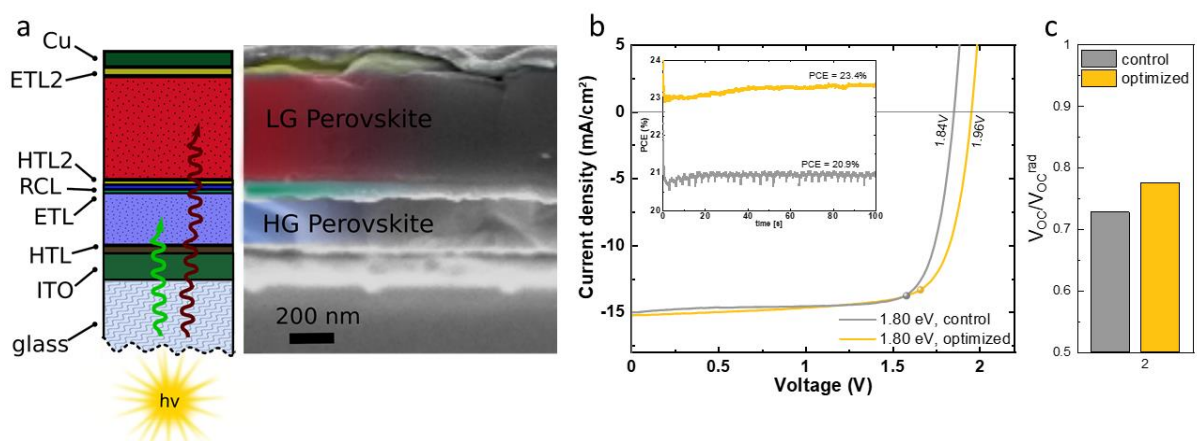

**Figure 3.** a) Schematic overview of the all-perovskite tandem structure alongside a cross-sectional scanning electron microscopy (SEM) image of an all-perovskite tandem with optimized HG perovskite. b) *JV* curves of a tandem with optimized 1.80 eV HG perovskite subcell, and a control tandem. In the inset, MMP tracking is displayed for both of these devices. The optimized wide-gap cell directly translates into improved tandem cell devices with efficiencies reaching over 23%. c) $V_{OC}/V_{OC}^{rad}$ ratios for the control and optimized tandems displayed in b.

In order to optimize the efficiency of the tandems further, we then implemented the three previously optimized high-gap perovskite compositions. The bandgap shift from 1.80 to 1.85 eV and 1.88 eV can be well seen in external quantum efficiency (EQE) spectra in all-perovskite tandems made thereof (**Figure 4a**). Naturally, the three different high gap subcells also influence the absorption onset of the LG perovskite. Using different bandgaps allowed us to improve the current matching between the HG and LG subcells, which is crucial for a monolithic tandem interconnection (see **Figure S3** for integrated EQE values and **Figure S18** for the current mismatch between HG and LG subcell).

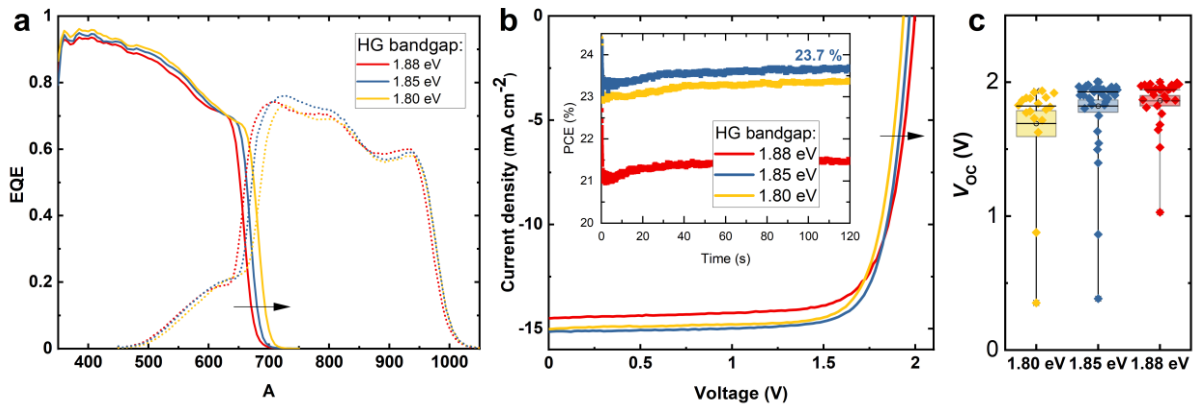

**Figure 4.** a) EQE spectra for all-perovskite tandems fabricated using three different HG perovskite bandgaps. These spectra alongside the integrated $J_{SC}$ values are shown in **Figure S3.** b) *JV* curves for representative tandems with three different HG perovskite bandgaps. Best performing MPP tracking is displayed in the inset. c) Statistical overview of the $V_{OC}$ for the different tandem systems. Further device statistics can be found in **Figure S4.**

As shown in **Figure 4b**, the $V_{OC}$ of fabricated all-perovskite tandems increases with increasing HG-perovskite bandgap, and best performing all-perovskite tandems based on a 1.85 eV HG-perovskite reach 23.7 % according to MPP tracking with a champion $V_{OC}$ of 2.00 V. The forward and reverse JV of this champion device can be found in **Figure S5.**

Interestingly, the $J_{SC}$ of our 1.80/1.27eV tandem combination exhibits a relatively high $J_{SC}$ equal to the integrated EQE from the HG subcell, although this tandem combination should be limited by the LG subcell producing a somewhat lower integrated EQE current. We performed all measurements with an illumination mask and confirmed that the spectral mismatch between our sun simulator and AM 1.5G is very small (see **Figure S17**) to exclude potential overestimations and thereby confirm that the $J_{SC}$ measured from *JV* is correct and not overestimated. Device statistics, displayed in **Figure S4**, further corroborate that our tandems – especially the 1.8/1.27 eV HG/LG combination - can operate without strict current matching. We believe this is caused by a rather low shunt resistance within the LG subcell, and show electrical simulations and sub-cell selective resistive photovoltage measurements highlighting the existence and the impact of low shunt resistances in the LG subcell on tandem solar cell operation and performance in the SI (see **Supplementary note 1** as well as **Figure S15**, **Figure S16**). Importantly, **Figure S16**, also shows that while the observed shunts in the LG cell can lift the current matching condition, the shunts will still reduce the PCE due to a concurrent reduction in *FF*, thus not causing an overestimation of the PCE. Indeed, looking at a statistical analysis of our fabricated devices we observe that the 1.80/1.27 eV tandem combination exhibits a larger $J_{SC}$ but lower *FF* in comparison to the better current matched 1.85/1.27eV and 1.88/1.27eV combinations (see **Figure S4)**.

**Sub-cell Analysis**

In order to get a deeper insight into the factors limiting the performance of these optimized tandem solar cells, we performed more detailed sub-cell selective EL measurements. Hereby we apply a forward bias to the tandem device that injects a current into both subcells. EL within both subcells then can be measured in the dark and easily distinguished by their corresponding photon energy, e.g. around 1.80, 1.85 and 1.88 eV for the HG-Perovskite subcells and 1.27 eV for the LG-subcell respectively, see also EL spectra displayed in **Figure S6**. In order to measure the EL quantum yield (ELQY) as a function of injection current, we used appropriate long-pass and short-pass filters together with a large-area Silicon-photodiode. Analogous to the PL, we can calculate the QFLS$_{EL}$ from the measured ELQY for each injection current $J_{\text{inj}}$ using equation 2.

$$\text{QFLS}_{EL} = k_B T \cdot \ln(\text{ELQY} \cdot \frac{J_{\text{inj}}}{J_{0,\text{rad}}}) \quad \text{(eq. 2)}$$

$J_{0,\text{rad}}$ values were calculated from EQE measurements, as detailed in the experimental methods, and we summarize results in **Figure S7** and **S8**, as well as **Table S1** for the different perovskites. Plotting the implied or pseudo-voltage ($_{ps}V_{EL}$ = QFLS$_{EL}$/e) on the x-Axis and the $J_{\text{inj}}$ current minus $J_{SC}$ ($J = J_{\text{inj}} - J_{SC}$) on the y-axis allows us to derive pseudo-light-$JV$ curves from the measured ELQY values. The derived pseudo-$JV$-characteristics are not only free of parasitic transport losses but most importantly reveal pseudo-$JV$-characteristics of the individual subcells, which cannot be accessed using standard $JV$ measurements under illumination. We show EL-pseudo-$JV$ curves of the individual subcells for all-perovskite tandems based on the 1.85/1.27 eV bandgap combination in **Figure 5a** alongside the summarized pseudo-$JV$ curves representing the resulting tandem as well as standard $JV$ characteristics under AM1.5G. Open and closed symbols refer to control and three-fold optimized HG subcells that we prepared within the same batch to avoid batch-to-batch variations. Comparison of the subcell pseudo-$JV$'s clearly shows that the improvement in the tandem $V_{OC}$ of $\Delta V_{OC}^{\text{tandem}}$~120 mV between control and optimized devices, directly results from the optimized HG subcell featuring an improvement of 120 mV. Both values are corroborated by the mean $V_{OC}$ improvement of 0.10 V we observe when evaluating statistics as shown as inset in **Figure 5a**.

To analyze this $V_{OC}$ improvement further, we compare in **Figure 5b** pseudo-$JV$ characteristics of the optimized tandem (closed circles) to the corresponding EL-pseudo-$JV$ characteristics of identically prepared single junctions (open square symbols). Notably, the LG perovskite pseudo-$JV$s in the tandem and in the single junction are identical, indicating that there are no $V_{OC}$ losses stemming from the integration in the tandem device. The optimized HG perovskite shows a slight (~20 mV) decrease in $V_{OC}$ upon incorporation in a tandem device. This could be caused by the processing of the LG perovskite on top of the HG subcell, but the difference is so small it could also be the result of device-to-device variation. On the other hand, the pseudo-$FF$s (pFF) and pseudo-PCEs (pPCE) of the HG subcell is improved in the tandem compared to pFF and pPCE of identically prepared HG single junctions. Our interconnect — comprising an ultrathin layer of indium oxide and a layer of tin oxide, both deposited by atomic layer deposition (ALD) on top of a spin coated layer of Al doped ZnO nanoparticles previously only applied in perovskite/organic tandems [37]— therefore can be considered quasi lossless.

Finally, we compare in **Figure 5b**, the pseudo-$JV$s (symbols) to regular $JV$ curves for HG and LG single junctions (lines) measured under AM1.5 and HG-filtered AM1.5G respectively. It can be seen here that the $V_{OC}$ and pseudo-$V_{OC}$ match very well. The *FF*s on the other hand are much lower than their corresponding pseudo-*FF*s, especially for the LG perovskite. This indicates that the cells suffer from severe transport losses, while the EL pseudo-*JV* measurements are only sensitive to the total non-radiative recombination losses analogously to a dark-*JV* curve and barely affected by resistive losses [40]. Overall, we can conclude that the *FF* can be improved from 74.6% to 84.6% by optimizing the charge transport in both subcells, which could enable an efficiency of 25.2%.

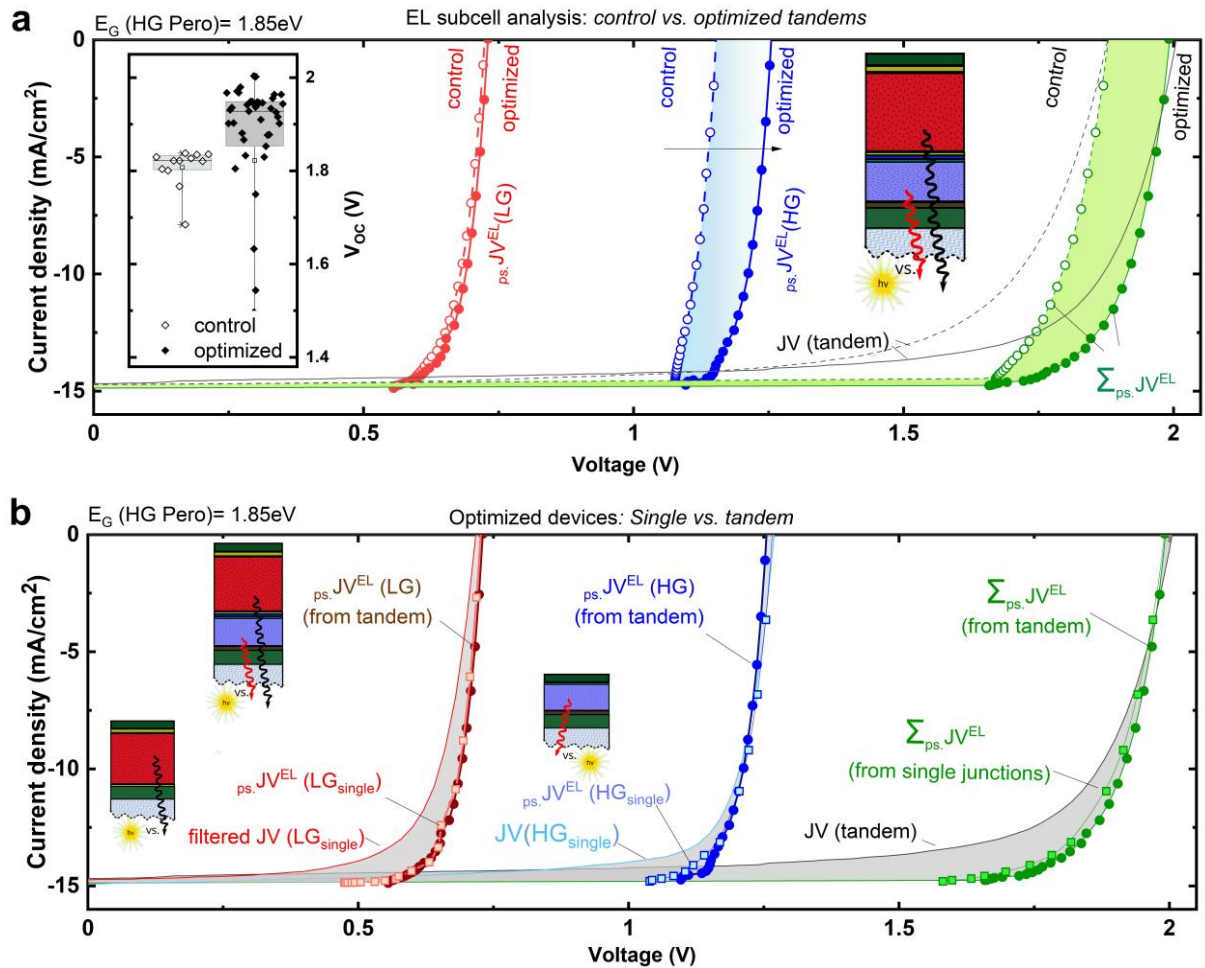

**Figure 5.** a) Pseudo *JV* curves from EL for an optimized and control tandem based on a 1.85 eV HG Cs$_{0.05}$(FA$_{0.55}$MA$_{0.45}$)$_{0.95}$Pb(I$_{0.55}$Br$_{0.45}$)$_3$ perovskite. Shown are the individual subcell pseudo-*JV* curves as well as the resulting (added) tandem pseudo-*JV* curves in comparison with traditional *JV* curves measured under AM1.5G. The inset displays $V_{OC}$s obtained from the *JV* curves of optimized and control tandems employing a 1.85 eV HG perovskite. b) Comparison between pseudo-*JV* curves obtained from an optimized 1.85/1.27eV tandem and pseudo-*JV* curves obtained from single junctions based on the same perovskites. Corresponding *JV* curves for both single junction and tandems are plotted alongside to highlight *FF* losses.

|  |  | $_{ps}V_{OC}$ [V] | $_{ps}FF$ [%] | $_{ps}PCE$ [%] |
|---|---|---|---|---|
| HG subcell | from EL | 1.26 | 88.2 | 16.4 |
| LG subcell | from EL | 0.73 | 80.6 | 8.8 |
| Tandem | from EL | 1.99 | 84.6 | 25.2 |
| Tandem | AM1.5 | 2.00 | 74.6 | 21.4 |
| HG single junction | from EL | 1.27 | 84.3 | 15.9 |
| HG single junction | AM1.5 | 1.27 | 78.7 | 15.6 |
| LG single junction | from EL | 0.73 | 81.3 | 8.8 |
| LG single junction | HG filtered AM1.5 | 0.72 | 71.2 | 7.6 |

**Table 2.** Summary of implied $V_{OC}$ and efficiency potentials from EL for optimized 1.85/1.27eV HG/LG based perovskite tandems, compared to device parameters measured under AM1.5G. Note that the comparison presented here was made on one exemplary device, while we summarize further device parameters and statistics in **Figure S3**. $J_{SC}$ in all cases was equal to $J_{gen}$ of 14.87 mA.

We further note that when using EL, a measurement that is performed in dark, care has to be taken on transient effects that are barely present under full AM1.5G illumination. Especially for the LG perovskite, the ELQY can increase upon light soaking of the cell as well as upon keeping the cell biased at $V_{OC}$ in the dark. We show examples of this effect in **Figure S9** and **S6** as well as the impact on extracted pseudo-$JV$ characteristics from EL measurements in **Figure S10**. Pseudo-JV characteristics we analyzed here were taken after the cell reached a steady-state comparable to standard $JV$ measurements under AM1.5G. We note that the derived $_{ps}V_{OC}$ at $J_{inj.} = J_{gen}$ conditions must equal the device $V_{OC}$ if Rau´s reciprocity is fulfilled [41]. This is generally observed for perovskite cells [42,43], and therefore a good sanity check of the EL and derived pseudo-$JV$ characteristics. If done properly, injection-dependent ELQY measurements reveal accurate pseudo-light $JV$ characteristics that allow us to obtain a comprehensive overview of the limiting factors.

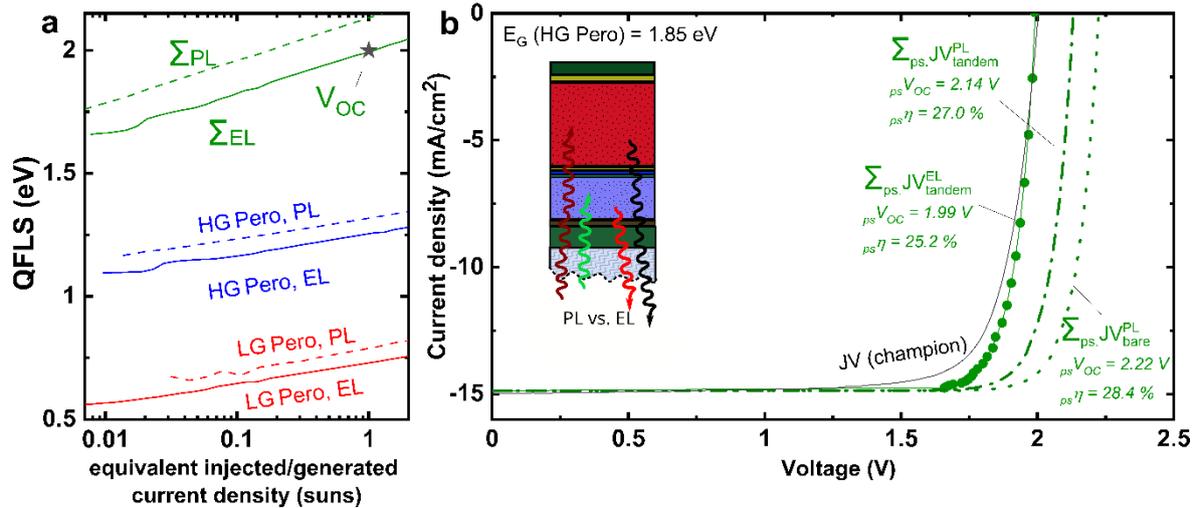

**Figure 6.** a) QFLS of the HG and LG perovskite subcells in a 1.85 eV/1.27eV tandem calculated from ELQY or PLQY as a function of the equivalent injected current density (ELQY) or generated current density (PLQY) respectively. Summed QFLS representing the actual tandem are further shown alongside the $V_{OC}$ obtained from *JV* characteristics under AM1.5G b) Pseudo-*JV*s reconstructed from EL (solid line) and PL (dashed line) tandem measurements, displayed alongside a pseudo-*JV* reconstructed from a PL measurement of isolated perovskite layers. The latter indicates the material efficiency potential.

Finally, we also perform intensity-dependent PL measurements on the optimized tandems, and compare the results to those from ELQY. **Figure 6a** displays the QFLS as a function of the equivalent injected current (ELQY) or generated current density (PLQY). If plotted on a semi-logarithmic scale the data follows a linear slope given by the subcell/tandem ideality factor. And while the ideality factors are relatively similar we notice significantly higher QFLS for PLQY measurements compared to ELQY for both the LG and the HG subcells. The discrepancy between these two values indicates energy level offsets present in the device stack, causing a difference between QFLS generated under illumination (i.e. from PLQY) and the device $V_{OC}$. Note that the QFLS from EL equals the device $V_{OC}$ if Rau´s reciprocity is fulfilled [41]. In **Figure S11**, simulated energy diagrams alongside a simulated *JV* curve corroborate the impact of potential energy level offsets. We note that the discrepancies between QFLS determined by EL and PL are also present in our single junction devices (see **Figure S12**), and thus stems from energy level offsets already present in the single junction stacks, rather than energy level offsets introduced by incorporation in the tandem cell or the recombination layer. Reducing such energy level offsets would enable us to minimize the QFLS discrepancy between ELQY and PLQY results, and push the efficiencies up further towards the potential indicated by the PLQY measurements at 27.0% versus 25.2% from ELQY. Corresponding intensity-dependent PL measurements, as well as a comparison between pseudo-*JV*s obtained from EL and PL measurements can be found in **Figure S13** and **Figure S14,** respectively.

Looking beyond the transport losses and QFLS – $V_{OC}$ mismatch, we ultimately also investigate the efficiency potential of the isolated absorber materials through intensity-dependent PLQY measurements. **Figure 6b** displays pseudo *JV* curves from ELQY and PLQY measurements on the tandems, alongside the pseudo-*JV* characteristics from PLQY measurements on the isolated perovskite layers, and clearly shows the limitations imposed by the transport layers. The combination of our 1.27 eV LG perovskite with the 1.85 eV HG perovskite reaches an absorber efficiency potential of 28.4% with an implied $V_{OC}$ of 2.22 V. Notably our three-fold optimized 1.85/1.27eV HG/LG champion tandem with a $V_{OC}$ of 2.00 V already reaches 84% of this material potential.

## Conclusion

In this work, we identified $V_{OC}$ losses in all-perovskite tandem solar cells and show that in our present material set, as well as many literature devices, nonradiative recombination within the high bandgap perovskite subcell dominates $V_{OC}$ and performance losses. We developed a multifaceted optimization route to improve the high bandgap subcell by replacing the HTL PTAA by 2PACz, adding oleylamine to the perovskite in combination with including a thin LiF layer between the perovskite and the ETL. In an additive manner, our combined approach enables high gap perovskite absorbers with high QFLS and an improved $V_{OC}$ potential, reaching 83% of their radiative $V_{OC}$ limit. The high $V_{OC}$ potential translated directly to the $V_{OC}$ of the all-perovskite tandems that were subsequently fabricated, and improved their steady-state power conversion efficiency to 23.7% for our champion combination of 1.85/1.27eV HG/LG perovskite subcells. We performed a thorough subcell analysis to disentangle further factors limiting the performance of these tandem devices and found that although there is still room for improvement of the $V_{OC}$s of both individual subcells, our ultra-thin InOx based interconnect is quasi lossless, and both subcells reach $V_{OC}$s equally high to those in their respective single junctions. The *FF* on the other hand is significantly lower than the pseudo-*FF* obtained from EL measurements, indicating significant transport losses. Reducing such transport losses would allow us to approach efficiencies of 25.2%, which is the efficiency potential for our 1.85/1.27eV HG/LG perovskite tandem combination extracted from EL-based pseudo-*JV* characteristics. We also observe a discrepancy between the *pseudo-$V_{OC}$* obtained from EL measurements, and the QFLS obtained from PLQY measurements, the latter being significantly higher. This indicates there are energy barriers present in the stack, which, when reduced, will provide a significant additional optimization potential, enabling efficiencies of up to 27.0%. Concluding, the indium oxide interconnect that was used in these tandems is quasi lossless, but both individual subcells, specifically the low-gap after the optimization of the high-gap perovskite, can still be improved to reach better performances. The insights of this extensive subcell-selective characterization provide crucial feedback and allow us to develop evidence-based optimization routes to improve the tandem efficiencies further in the future.

**Author contributions**

Jarla Thiesbrummel: conceptualization, formal analysis, investigation, visualization, writing–original draft; Francisco Peña-Camargo: conceptualization, formal analysis, investigation, visualization; Kai O. Brinkmann: investigation, formal analysis, visualization, writing–review and editing; Emilio Gutierrez-Partida: investigation; Fengjiu Yang: investigation, writing–review and editing; Jonathan H. Warby: writing–review and editing; Steve Albrecht: funding acquisition, resources, supervision; Thomas Riedl: funding acquisition, writing–review and editing, supervision; Dieter Neher: funding acquisition, resources, writing – review and editing, supervision; Henry J. Snaith: funding acquisition, writing – review and editing, supervision; Martin Stolterfoht: conceptualization, funding acquisition, writing – review and editing, supervision; Felix Lang: conceptualization, formal analysis, investigation, visualization, writing-original draft, supervision

**Conflict of interest**

H.J.S. is co-founder and CSO of Oxford PV Ltd.

**Acknowledgements**

This work was part-funded by EPSRC, project number EP/S004947/1. We acknowledge funding from the Deutsche Forschungsgemeinschaft (DFG, German Research Foundation) within the SPP 2196 (HIPSTER 424709669 and SURPRISE 423749265). We further acknowledge financial support by the Federal Ministry for Economic Affairs and Energy within the framework of the 7th Energy Research Programme (P3T-HOPE, 03EE1017C) and HyPerCells (a joint graduate school of the Potsdam University and the Helmholtz Zentrum Berlin). J.T. thanks the Rank Prize fund for financial support. M.S. acknowledges the Heisenberg program from the Deutsche Forschungsgemeinschaft (DFG, German Research Foundation) for funding – project number 498155101. F.L. acknowledges financial support from the Alexander Von Humboldt Foundation via the Feodor Lynen program.

# Supporting Information
**Supplementary Methods**

**Device fabrication of *pin*-type cells:**

*Substrates and HTL:*
Pre-patterned (2.5×2.5 cm$^2$, 15 Ω/sq) ITO substrates (Psiotec, UK) were sonicated for 10 minutes subsequently in acetone, 3% Hellmanex solution in deionized (DI) water, DI-water and isopropanol in order to clean them. After an oxygen plasma treatment (4 min, 120 W), the substrates for the mixed-metal lead-tin perovskite devices were transferred to a laminar flow bench, while the other substrates were transferred to a N$_2$-filled glovebox. When the HTL is 2PACz, the substrates are transferred to an ozone treatment for 30 min after the cleaning process and they do not undergo any plasma treatment. Afterwards, they are transferred, as fast as possible, to the N$_2$-filled glovebox.

For the mixed-metal *pin*-type cells shown in the main text, a PEDOT:PSS (poly(3,4-ethylenedioxythiophene) polystyrene sulfonate) (Heraeus, AL3083) layer was spin-coated in air from a 1:2 (PEDOT:PSS):methanol solution, that had been filtered using a 0.45 µm GMF filter. The PEDOT:PSS/methanol solution was spin-coated at 4000 rpm for 45 seconds at a ramp of 1500 rpm/s After 10 minutes annealing on a pre-heated hotplate at 120 °C, the films were transferred immediately to N$_2$ filled glovebox.

For the *pin*-type Pb-based cells shown in the main text, 60 µL of PTAA (Poly-[bis-(4-phenyl)-(2,4,6-trimethylphenyl)-amin]) solution (Sigma-Aldrich, 1.75 mg/mL in toluene) was spin-coated at 6000 rpm for 30 seconds with a ramp of 2000 rpm/s. After 10 min annealing on a hotplate at 100 °C, the films were cooled down to room temperature. The estimated thickness of the PTAA layer is 8 nm. Thereafter, 60 µL of PFN-Br (Poly(9,9-bis(3'-(N,N-dimethyl)-N-ethylammoinium-propyl-2,7-fluorene)-alt-2,7-(9,9-dioctylfluorene))dibromide) solution (1-Material, 0.5 mg/ml in methanol) was deposited on top of the PTAA layer dynamically at 4000 rpm for 30 s resulting in a film with thickness below the detection limit of our AFM (< 5 nm). No further annealing of the HTL took place after this.

Alternatively, a previously sonicated solution of 2PACz (TCI, 1.0 mg/mL, in ethanol) was spin-coated at 3000 rpm for 30 seconds with a ramp of 3000 rpm/s. The substrates were annealed at 100 °C for 10 minutes and cooled down to room temperature before the deposition of the perovskite layer.

*Perovskite solutions:*
The 1.2 M mixed-metal lead-tin perovskite solutions were prepared by dissolving FAI, CsI, SnI$_2$ and PbI$_2$, together with, relative to their respective metals, 6 molar % Pb(SCN)$_2$ and 10 molar & SnF$_2$ in a 4:1 DMF:DMSO mixture. We note that all precursors used for this solution were stored and weighed in a N$_2$-filled glovebox, to prevent contamination of the solution with O$_2$ or H$_2$O. The solution was stirred for 2 hours at room temperature. Finally, the solution was filtered using a 0.45 µm PTFE filter.

The solutions for the lead-based perovskites layers were prepared as follows: 1.2 M FAPbI$_3$ solution was prepared by dissolving FAI and PbI$_2$ in DMF:DMSO (4:1 volume ratio) which contains a 10%-molar excess of PbI$_2$. The 1.2 M MAPbBr$_3$ solution was made by dissolving MABr and PbBr$_2$ in DMF:DMSO (4:1 volume ratio) which contains a 10 %-molar excess of PbBr$_2$. The solutions were stirred overnight at room temperature. By mixing these FAPbI$_3$ and MAPbBr$_3$ solutions in a ratio of 60:40, 55:45 and 50:50 respectively, we get what we call "MAFA" solutions. Lastly, 42 µL of a 1.5 M CsI solution in DMSO was mixed with 958 µL of each one of the MAFA solutions resulting in nominal triple cation perovskite stoichiometries of Cs$_{0.05}$(FA$_{0.60}$MA$_{0.40}$)$_{0.95}$Pb(I$_{0.60}$Br$_{0.40}$)$_3$ , Cs$_{0.05}$(FA$_{0.55}$MA$_{0.45}$)$_{0.95}$Pb(I$_{0.55}$Br$_{0.45}$)$_3$

and $Cs_{0.05}(FA_{0.50}MA_{0.50})_{0.95}Pb(I_{0.50}Br_{0.50})_3$. These three triple cation perovskites have bandgaps of 1.80, 1.85, 1.88 eV respectively. Optimized recipes included a trace amount (0.002 molar %) of oleylamine (Sigma Aldrich, technical grade, 70%) that was added directly to the final perovskite solutions.

*Perovskite film fabrication*:

The mixed-metal $FA_{0.83}Cs_{0.17}Pb_{0.5}Sn_{0.5}I_3$ perovskite films were deposited by spin-coating 120 μL perovskite solution at 3000 rpm for 45 s with a ramp of 1000 rpm/s. 25 s after the start of the spinning process, the spinning substrate was washed with 200 μL anisole, which was deposited in the centre of the film. By the end of the spinning process, the perovskite films turned dark brown. The perovskite films were then annealed at 100 °C for 10 minutes on a preheated hotplate inside the $N_2$ filled glovebox. During the annealing process, the perovskite films turned black.

All triple cation perovskite films were prepared by depositing 120 μL perovskite solution and spin-coating at 4000 rpm for 40 s at a ramp of 1334 rpm/s. 10 s after the start of the spinning process, the spinning substrate was washed with 300 μL ethylacetate for approximately 1 s (the anti-solvent was placed in the center of the film). We note, that by the end of the spinning process the perovskite film turned dark brown. The perovskite film was then annealed at 100 °C for 1 h on a preheated hotplate where the film turned slightly darker.

*ETL and Top Contact*:

After annealing, the samples were transferred to an evaporation chamber where fullerene $C_{60}$ (30 nm), 2,9-Dimethyl-4,7-diphenyl-1,10-phenanthroline BCP (8 nm) and copper (100 nm) were deposited under vacuum ($p = 10^{-7}$ mbar). The overlap of the copper and the ITO electrodes defined the active area of the pixel (6 mm$^2$). For optimized HG perovskites, (0.8 nm) of LiF (Alfa Aesar) was deposited by thermal evaporation before the $C_{60}$.

**Device fabrication of tandems:**

To fabricate tandem devices, the high gap perovskite cells were prepared as described above, up until the deposition of the C60 layer. Hereafter, aluminum zinc oxide (AZO) nanoparticle dispersion (N21X purchased from Avantama) was diluted 1:2 ratio with isopropanol and spin-coated at 4000 rpm for 20 s with a ramp of 6 s, followed by a 90 min annealing step at 80 °C.

Thereafter the samples were transferred into our Beneq TFS-200 system without inert break. $SnO_x$ and $InO_x$ layers were sequentially grown from tetrakis(dimethylamino)tin(IV) (TDMA-Sn, Strem) + water and Cyclopentadienylindium (CpIn, Strem), oxygen (purity 99.999%) + water respectively. TDMA-Sn and CpIn were used from hot sources kept at 45 °C, and 50 °C respectively. Water was kept in a liquid source at room temperature. Reactor temperature during both deposition processes was set to 80 °C.

After the deposition of the interconnect, the tandems were finished in the same manner as described above for the low-gap perovskite single junctions, by deposition of PEDOT:PSS, low-gap perovskite and ETL & top contact.

**Current density-voltage characteristics:** $JV$-curves were obtained in a 2-wire source-sense configuration with a Keithley 2400 inside a nitrogen filled glovebox. An Oriel class AAA Xenon lamp-based sun simulator was used for illumination providing approximately 100 mW cm$^{-2}$ of AM1.5G irradiation and the intensity was monitored simultaneously with a Si photodiode. The spectrum of the sun simulator was measured and compared to the AM1.5G spectrum (**Figure S17**). Spectral mismatch factors were calculated for devices with different bandgaps, and reported alongside the spectrum in **Figure S17**. The exact illumination intensity was used for efficiency calculations, and the simulator was calibrated with a KG3 filtered silicon solar cell (certified by Fraunhofer ISE). For measurements on the low-bandgap devices, the calibration was cross-checked with a HOQ filtered silicon solar cell (certified by Fraunhofer ISE). The obtained short-circuit current density ($J_{SC}$) is checked by integrating the product of the External Quantum Efficiency and the solar spectrum, as discussed in **Supplementary Note 1**, and matched within 2%. The temperature of the cell was fixed to 25 °C and a voltage ramp (scan rate) of 67 mV/s was used. To precisely define the active area, all measurements presented here were taken with a shadow mask, with an area of 0.0432 cm$^2$.

**External Quantum Efficiency:** External quantum efficiencies were measured inside a nitrogen-filled glovebox as a function of wavelength from 300 nm to 1200 nm with a step of 5 nm using a custom-built small spot EQE system. For measurements of the perovskite top cell, infrared bias light was applied while for measurements of the perovskite bottom cell, blue bias light was applied using appropriate LEDs.

**Electroluminescence measurements (EL):** Electroluminescence measurements were obtained by applying a bias voltage to the cell, and recording the electroluminescence emitted by the cell with a photodiode. To distinguish between the emission of the different subcells, short- and longpass filters were used. Calibration measurements were performed to account for the extra distance introduced by the filters between the cell and the photodiode. These electroluminescence measurements were performed on unencapsulated cells in a N$_2$ filled glovebox. To account for the spectral response of the photodiode, the EL spectra were recorded using an Andor Solis setup with a Si detector. A bias voltage was applied to the encapsulated cell, after which the spectrum was recorded. All measurements were repeated several times to check for consistency and changes over time.

**Photoluminescence measurements (PL):** Photoluminescence measurements were obtained by illuminating an encapsulated device or perovskite comprising stack with a 520 nm CW laser (for HG perovskites) or a 808 nm CW laser (for LG perovskites). The resulting PL spectra were measured using an Andor Solis setup with a Si detector. All measurements were repeated several times to check for consistency. Intensity-dependent photoluminescence measurements were performed similarly, measuring the PL for different light intensities (in an increasing or decreasing manner). The applied light intensities were monitored with a Si photodiode.

**Rising Photovoltage measurements (RPV):** Rising photovoltage measurements were obtained for both subcells of the tandem by applying a short laser pulse (IR to measure the low gap subcell, green to measure the high gap subcell), after which the transient photovoltage is recorded across a large (1MΩ) resistance. This large resistance is used to create a large RC time, which enables visualisation of the transit time of the carriers through the tandem device. Single junctions were also measured for comparison – using the same laser wavelength as was used to measure their corresponding subcells. The measurements were repeated for 3 different laser fluences.

**Numerical drift-diffusion simulations:** The simulations were performed using Setfos 5.1. Details of the parameters used can be found in **Table S2.**

**Electric circuit simulations:** The simulations were performed using LTspice electric circuit simulation software.

**Detailed Balance Calculations:** Detailed balance calculations of all-perovskite tandem solar cells were performed assuming a step function absorption profile for each bandgap using a freely available python code, ref [1]. The temperature was set to 300K. Calculations did not include a spectrally selective reflector between the subcells.

**QFLS$_{rad}$ Calculations:** The quasi-Fermi level splitting in the individual sub-cells was calculated following detailed balance, linking the radiative recombination density of free charges ($J_{rad}$) with the chemical potential per free electron-hole pair ($\mu$) or the quasi-Fermi level splitting (QFLS) in the respective active material.[2,3]

$$J_{rad} = J_{0,rad} \exp{(\text{QFLS}/k_B T)}, \qquad (\text{eq. S1})$$

Here, $J_{0,rad}$ is the radiative thermal recombination current density in the dark, $k_B$ the Boltzmann constant and $T$ the temperature. Equation S1 is a simplification of Würfel's generalized Planck law, and thereby only valid for a QFLS that is a few $k_B T$ smaller that the bandgap $\mu < E_G - 3k_B T$.[4] If radiative recombination comes only from free charges, the radiative recombination current is identical to the photoluminescence yield times the elementary charge, that is $J_{rad} = \phi_{PL} \cdot e$. Moreover, we can define the photoluminescence quantum yield (PLQY) as the ratio of radiative to total recombination ($J_{R,tot}$), where the latter is identical to the generation current density ($J_G$) under open-circuit conditions ($V_{OC}$)

$$\text{PLQY} = \frac{J_{rad}}{J_{R,tot}} = \frac{J_{rad}}{J_G} \qquad (\text{eq. S2})$$

The QFLS is then given by

$$\text{QFLS} = k_B T \ln{(\text{PLQY} * \frac{J_G}{J_{0,rad}})} \qquad (\text{eq. S3})$$

With T = 300K and the measured PLQY values. For a PLQY = 1, we further get the radiative limit of the QFLS (QFLS$_{rad}$) via:

$$\text{QFLS}_{rad} = k_B T \ln{(\frac{J_G}{J_{0,rad}})} \qquad (\text{eq. S4})$$

We note that equations 2 and 4 are only valid if the spectral dependence of $J_{rad}$ is identical to $J_{0,rad}$, meaning recombination goes through the same channels regardless of the QFLS. The generation current density $J_G$ was approximated with the short-circuit current density of the complete solar cell. The $J_{0,rad}$ was estimated by integrating the overlap of the external quantum efficiency (EQE) of the respective subcell with the black body spectrum $\phi_{BB}$ at 300 K over the energy.

$$J_{0,rad} = \int \text{EQE} \, \phi_{BB} \, d\epsilon \qquad (\text{eq. S5})$$

with

$$\phi_{BB} = \frac{1}{4^2 \hbar^3 c^2} \cdot \frac{E^2}{\exp{\left(\frac{E}{k_B T}\right)} - 1} \qquad (\text{eq. S6})$$

Results are summarized in Figure S7, Figure S8 and Table S1.

**Supplementary Figures**

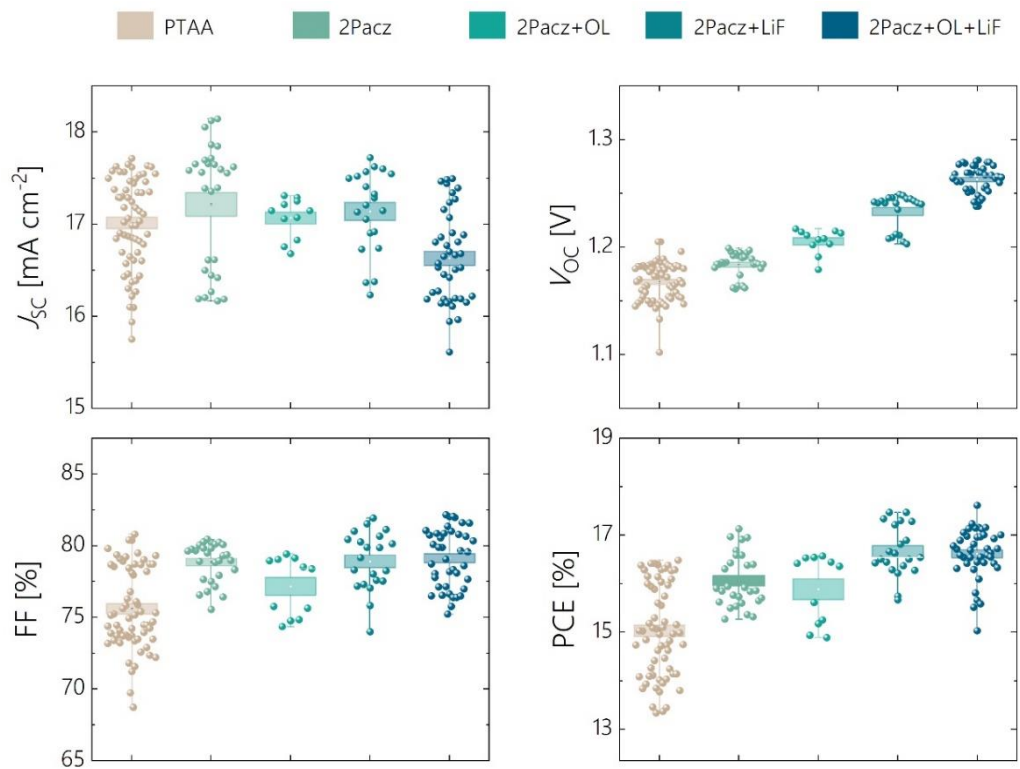

**Figure S1.** Device statistics for the different optimization steps, for the 1.80 eV HG perovskite

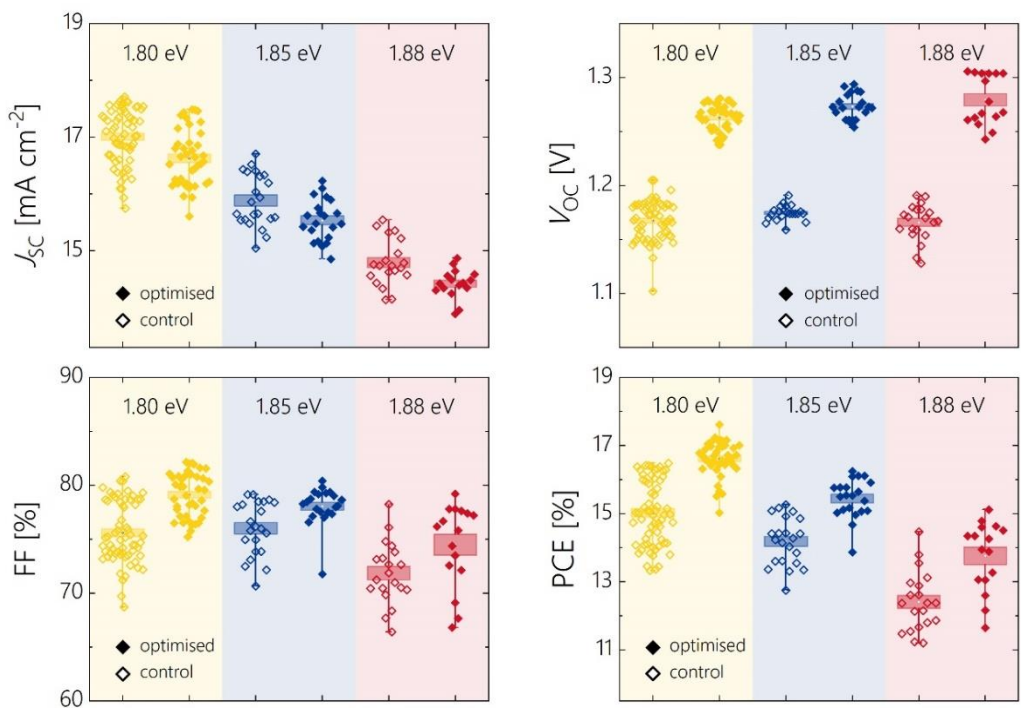

**Figure S2.** Device statistics for control (open symbol) and fully optimized (closed symbol) single junction perovskite solar cells with bandgaps of 1.80, 1.85, and 1.88 eV

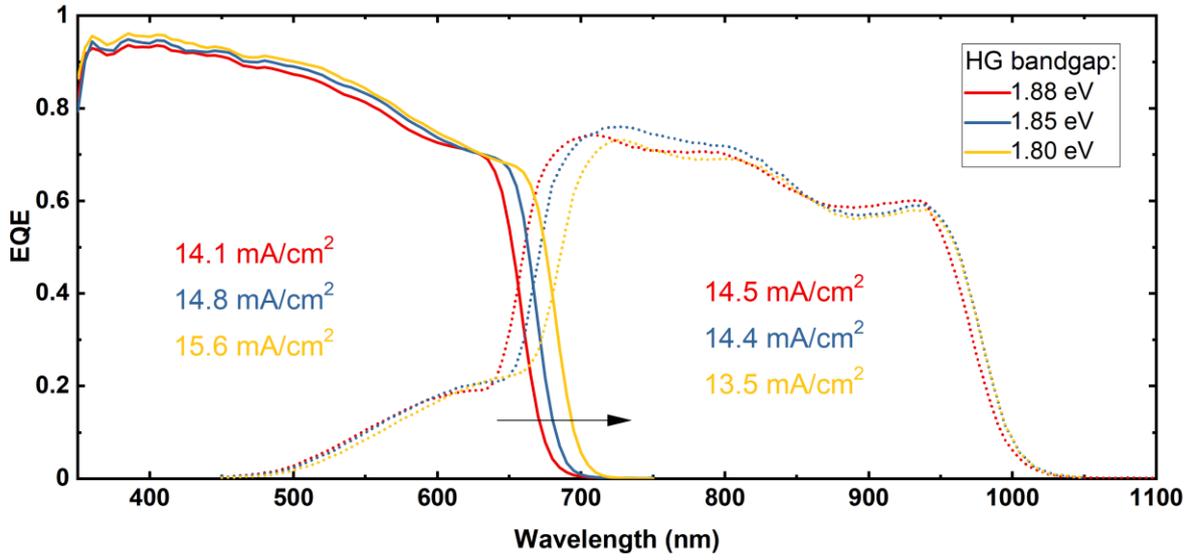

**Figure S3**. EQE and integrated currents for tandems with three different HG perovskite bandgaps.

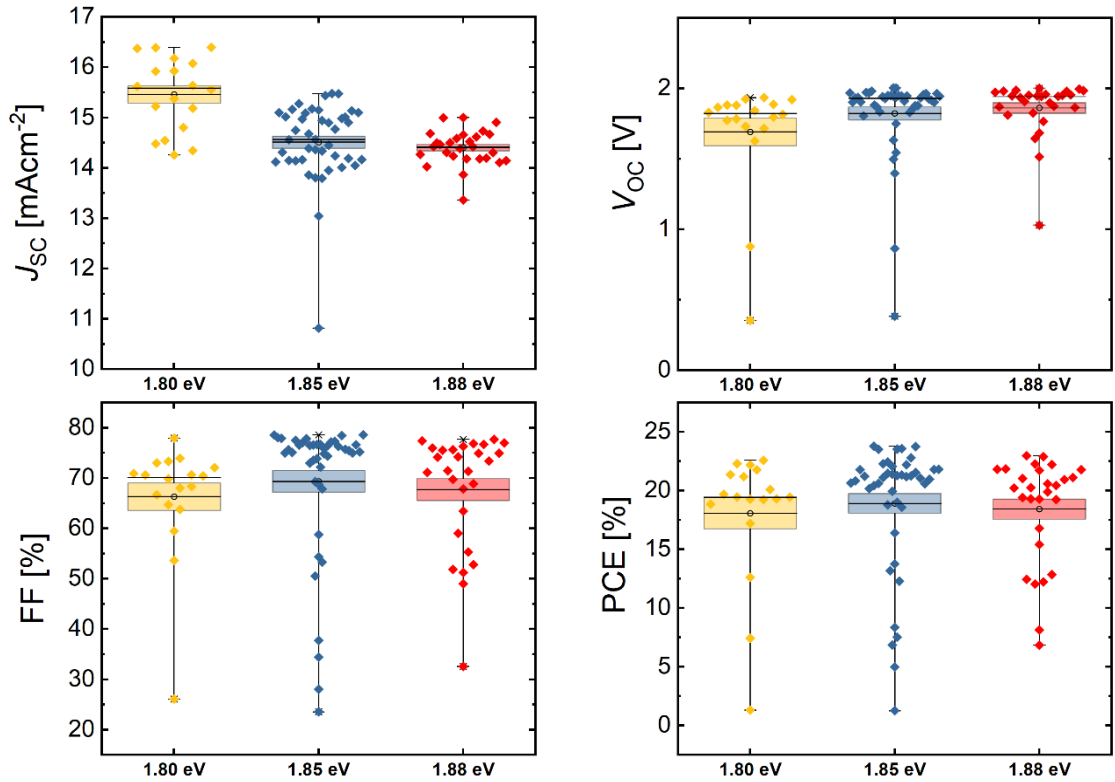

**Figure S4.** Cell statistics for the optimized all-perovskite tandem solar cells based on 1.80, 1.85 and 1.88 eV high gap perovskites.

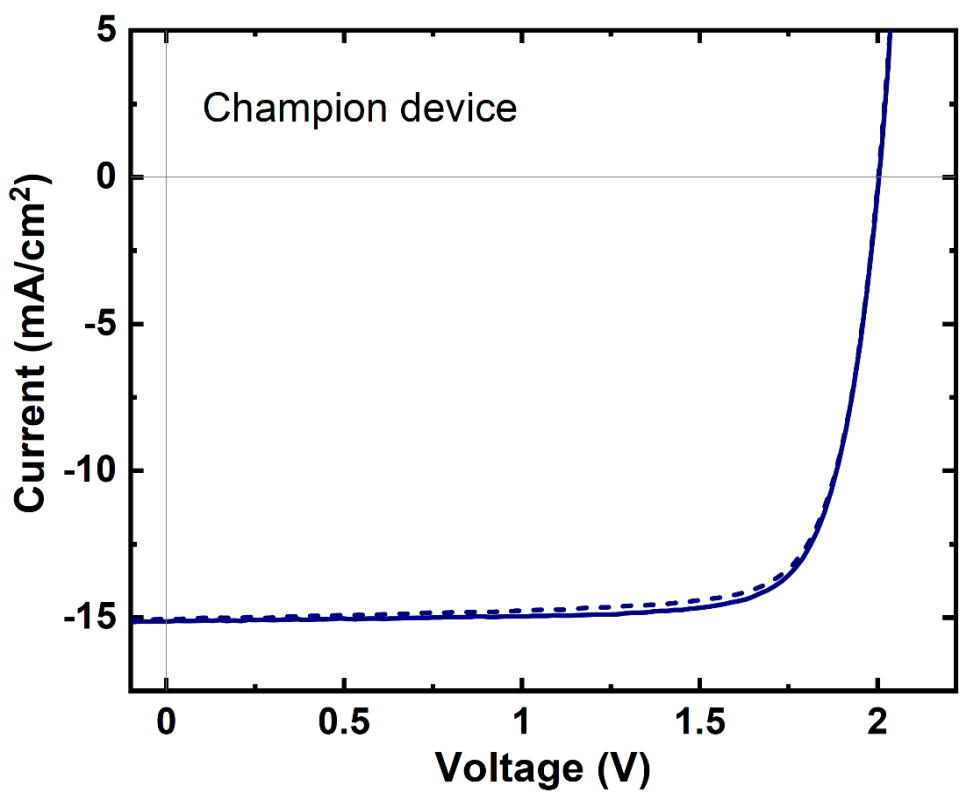

**Figure S5.** Forward and reverse JV scan of the champion device, based on a 1.85 eV HG perovskite in combination with a 1.27 eV LG perovskite. Device parameters are: $J_{sc}$ = 15.1 mA/cm², $V_{oc}$ = 2.00 V, $FF$ = 78.6 %, $PCE$ = 23.8 % (reverse) and $J_{sc}$ = 15.0 mA/cm², $V_{oc}$ = 2.00 V, $FF$ = 77.8 %, $PCE$ = 23.4 % (forward).

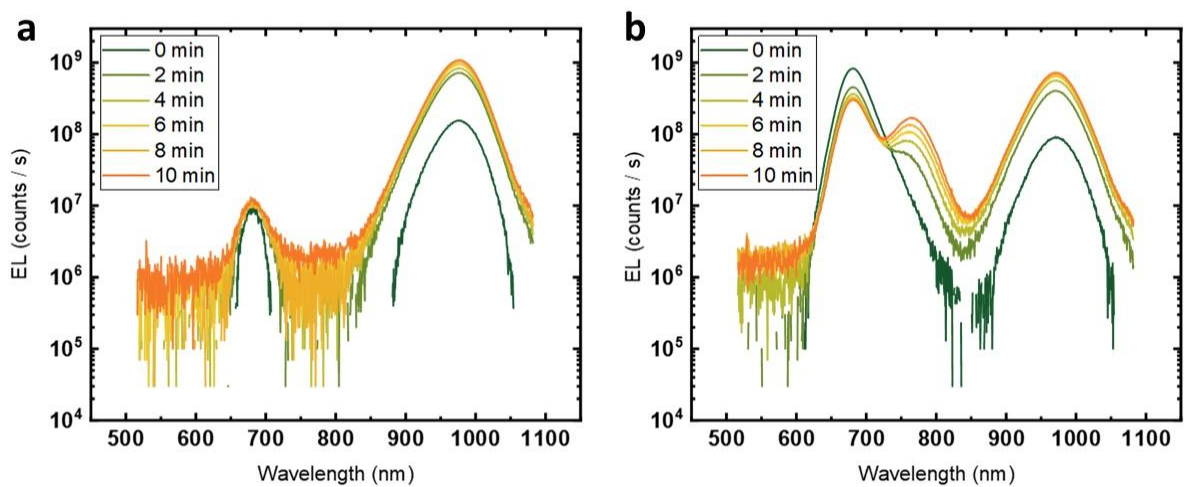

**Figure S6.** EL spectra while continuously applying $V_{OC}$ to the cell for a) a tandem with the control HG perovskite and b) a tandem with the optimized HG perovskite. The optimized HG perovskite displays an increased EL, but also a stronger halide segregation. The emission from the segregated phase was not considered in the determination of the QFLS. The reasons for this were recently discussed in ref. [5]

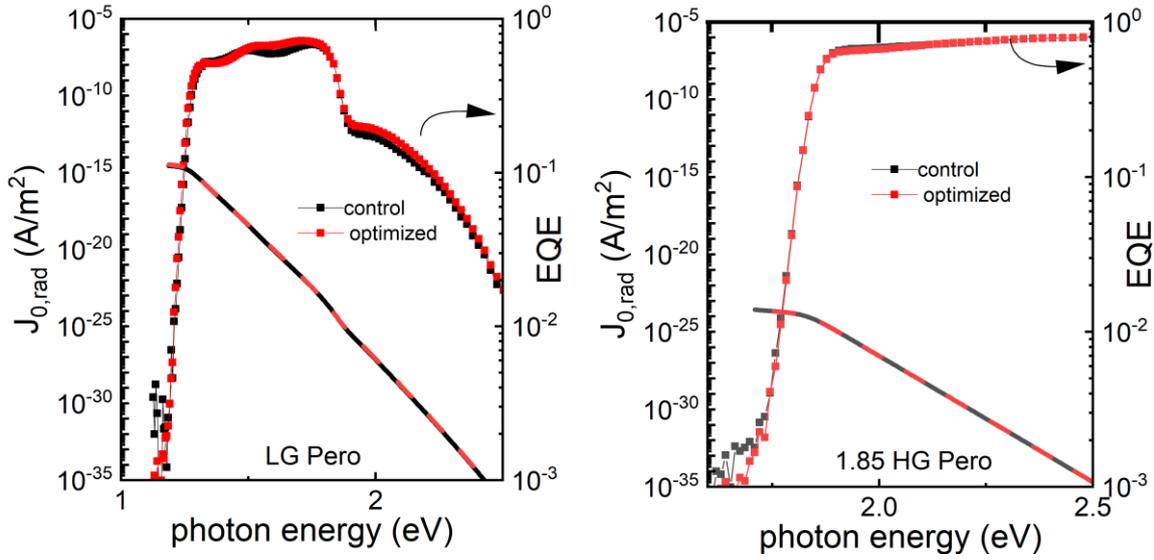

**Figure S7.** calculations of the radiative dark recombination current $J_{0,rad}$, from EQE measurements for the low gap (left) and high gap (right, 1.85 eV) perovskites. Calculation details are given in the experimental methods above, see equation S5 and S6.

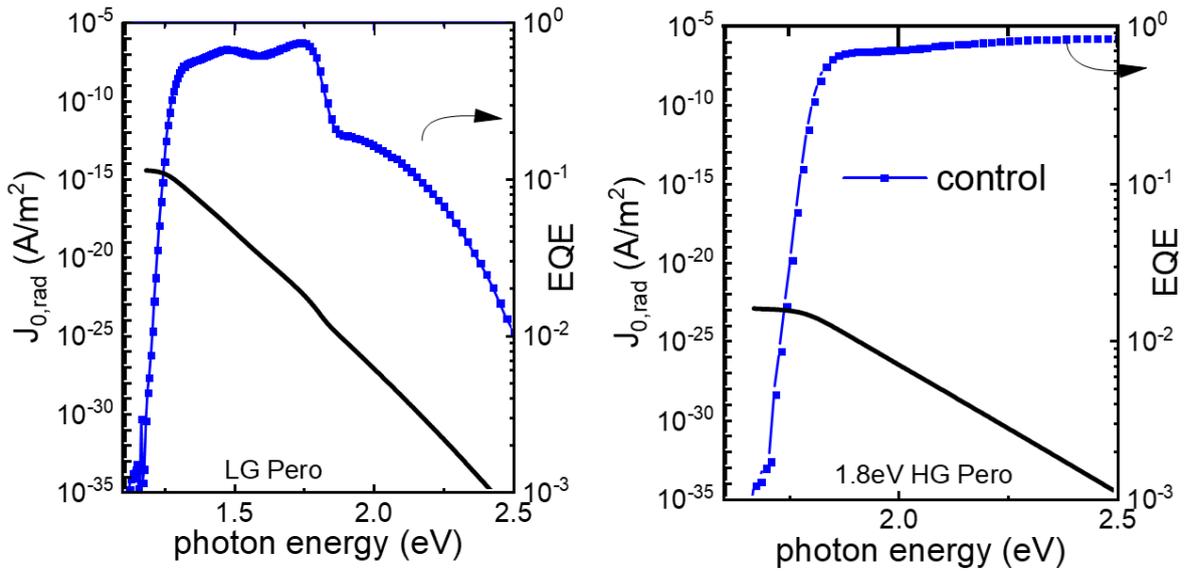

**Figure S8.** calculations of the radiative dark recombination current $J_{0,rad}$, from EQE measurements for the low gap (left) and high gap (right, 1.80 eV) perovskites. Calculation details are given in the experimental methods above, see equation S5 and S6.

|                       | HG Pero   | LG Pero   |
|-----------------------|-----------|-----------|
| 1.27/1.85eV optimized | 2.64E-24  | 3.37E-15  |
| 1.27/1.85eV control   | 2.76E-24  | 2.86E-15  |
| 1.27/1.80eV control   | 8.90E-24  | 3.08E-15  |

**Table S1.** Values of the radiative dark recombination current J0rad, as calculated in **Figure S7** and **Figure S8**. Calculation details are given in the experimental methods above, see equation S5 and S6.

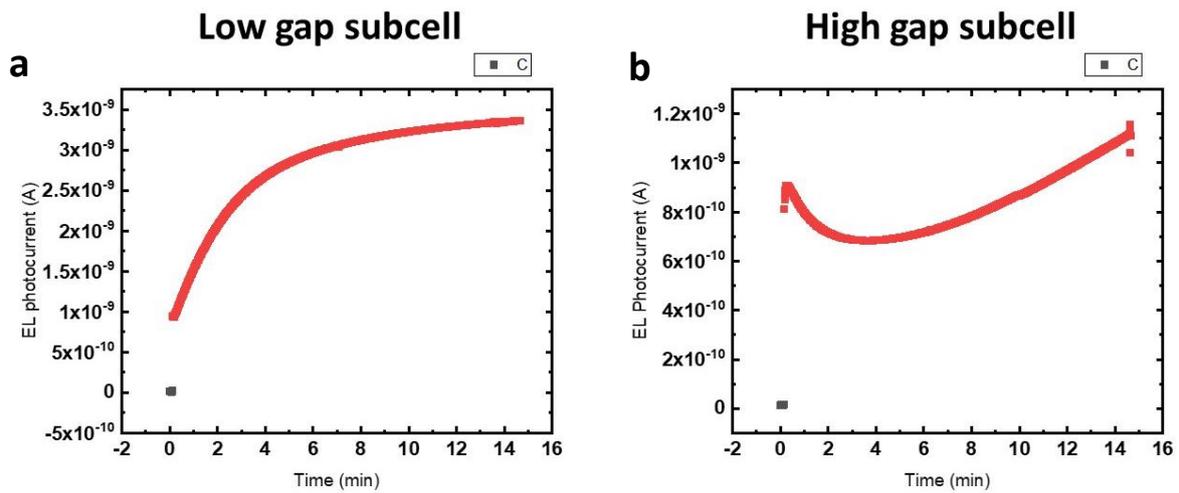

**Figure S9.** EL photocurrent as a function of time while applying $V_{OC}$ to the cell for both the LG (a) and the HG (b) subcell.

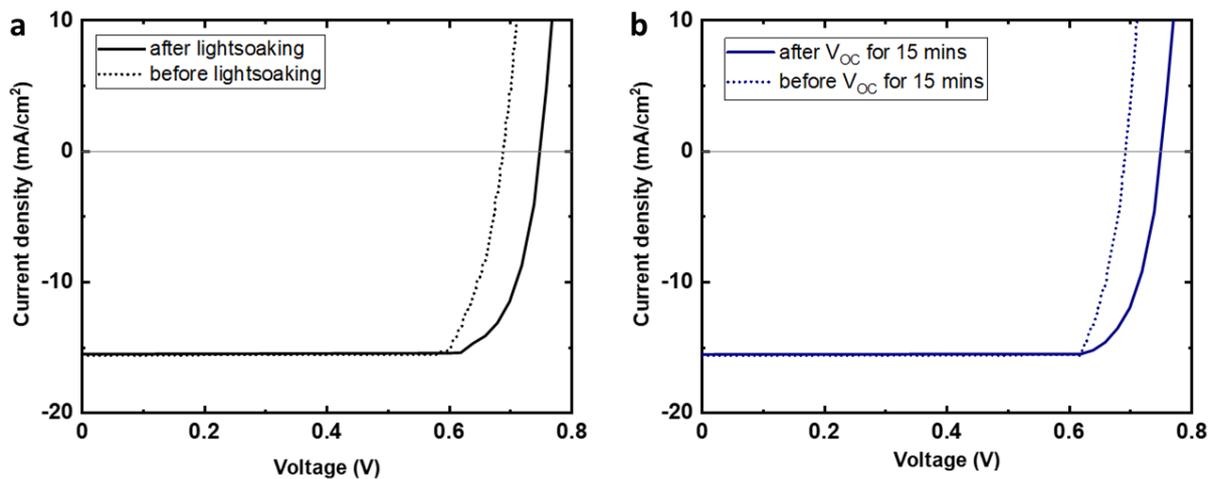

**Figure S10.** a) Pseudo-*JV*s for the LG subcell obtained from EL before and after light soaking, as well as b) pseudo-*JV*s obtained before and after keeping the cell at $V_{OC}$ in the dark for 15 minutes.

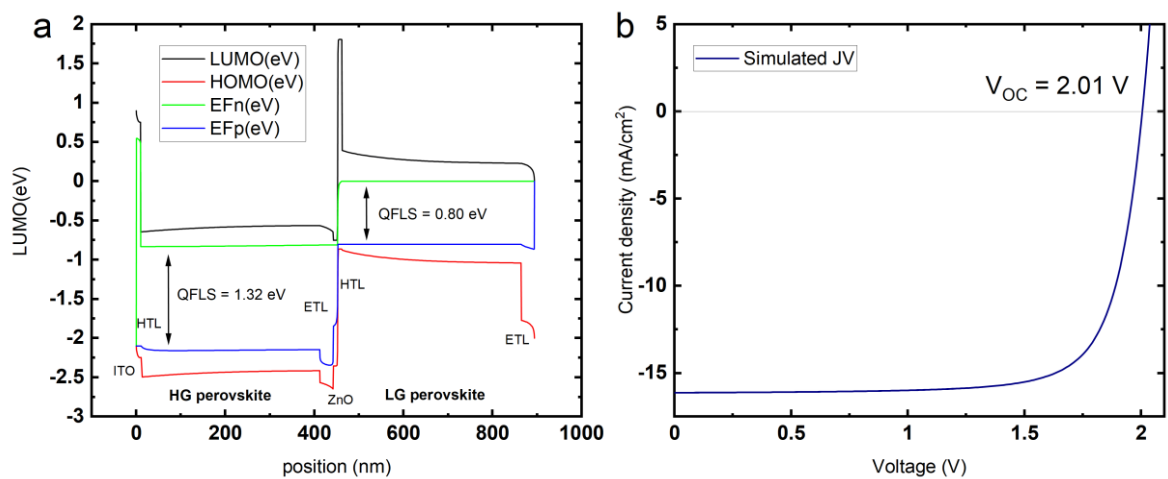

**Figure S11.** Simulated band structure (a) and *JV* curve (b) for an all perovskite tandem with bandgaps of 1.27 eV and 1.85 eV for the LG and HG, respectively. It can be seen that the QFLS of the individual subcells add up to 2.12 V, while the V$_{OC}$ of the tandem is limited to 2.01 V. In the simulations, this was reproduced by implementing an energy level offset at the HTL/perovskite interfaces, although we note that we do not exclude that there are other factors that lead to a QFLS-V$_{OC}$ mismatch [6,7]. Simulations were carried out using SETFOS 5.2.5., with parameters summarized in Table S2.

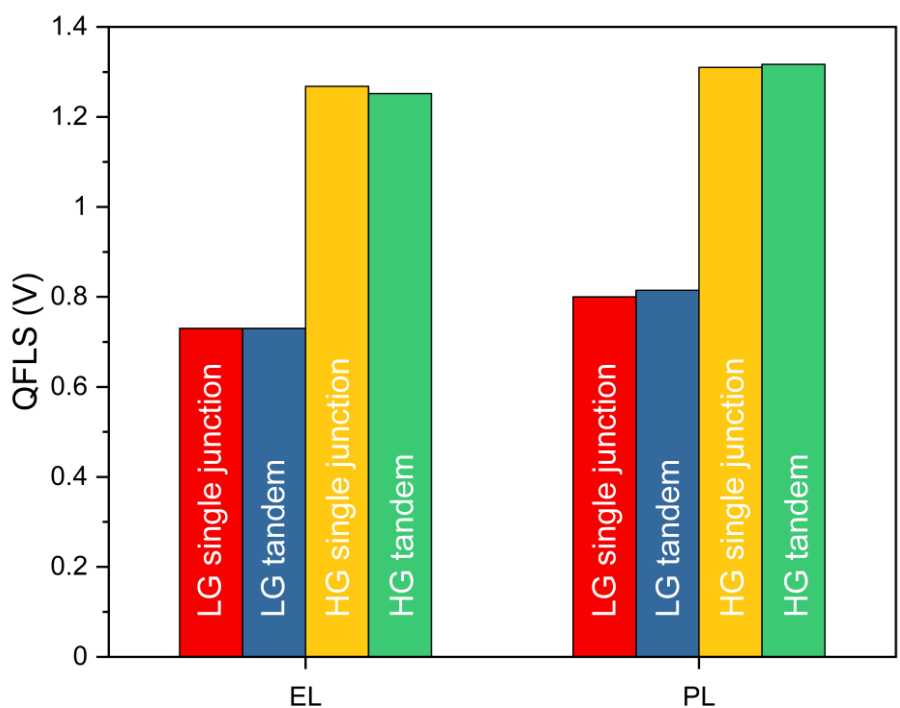

**Figure S12.** QFLS determined from EL (left) and PL (right) for single junctions as well as tandem subcells.

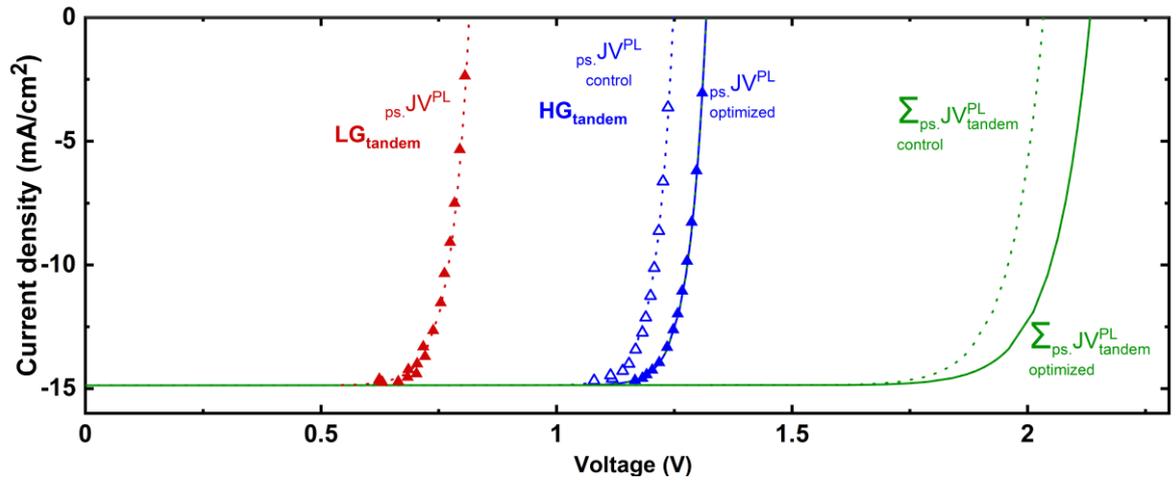

**Figure S13.** pseudo-*JV*s obtained from PL measurements for the optimized and control tandems

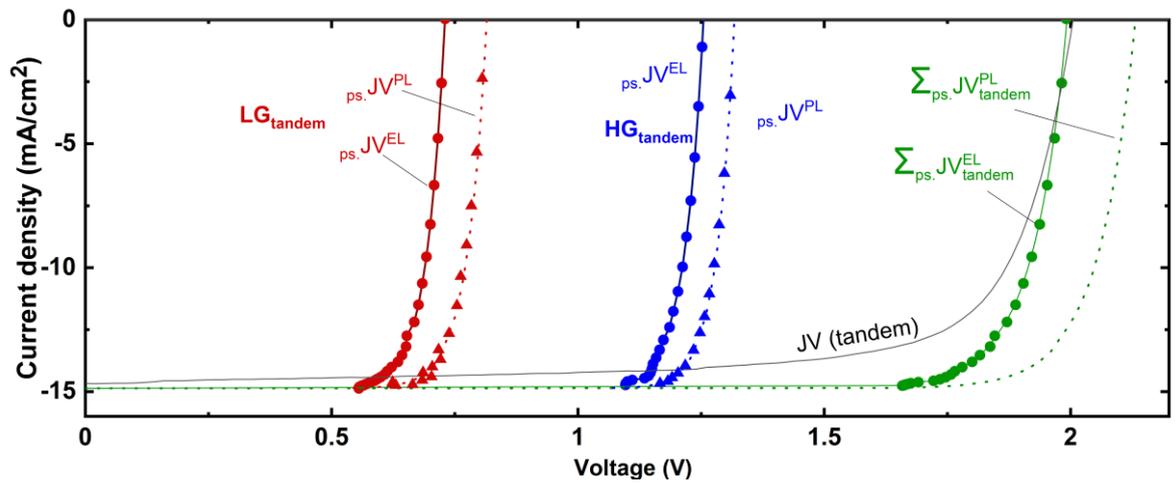

**Figure S14.** Comparison between pseudo-*JV* curves obtained from iPLQY and EL measurements, alongside *JV* curves of the tandem and corresponding single junctions.

**Table S2.** Setfos simulation parameters as used for all-perovskite tandem simulations displayed in **Figure S11**. These parameters are largely based on those used in our previous work [8,9]. In order to simulate the interfaces of the perovskite with the transport layers, a thin (1 nm) perovskite interface layer was added on both sides of the bulk with a higher concentration of traps.

| Parameter | Thickness | HOMO | $N_0$ | LUMO | $N_0$ | Dielectric constant | electron mobility | hole mobility | Electron trap density |
|---|---|---|---|---|---|---|---|---|---|
| Unit | nm | eV | $m^{-3}$ | eV | $m^{-3}$ | | $cm^2/Vs$ | $cm^2/Vs$ | $cm^{-3}$ |
| **Layers** | | | | | | | | | |
| **ITO** | 120 | | | | | | | | |
| **PTAA** | 10 | 5.5 | 1.0E+26 | 2.5 | 1.0E+26 | 3.5 | 1.0E-06 | 3E-03 | |
| **HTL/Perovskite interface** | 1 | 5.5 | 2.2E+24 | 3.9 | 2.2E+24 | 22 | 1.0 | 1.0 | |
| **Perovskite** | 400 | 5.75 | 2.2E+24 | 3.9 | 2.2E+24 | 22 | 1.0 | 1.0 | 2.0E+15 |
| **ETL/Perovskite interface** | 1 | 5.75 | 2.2E+24 | 3.9 | 2.2E+24 | 22 | 1.0 | 1.0 | 5.0E+17 |
| **C60** | 30 | 5.9 | 1.0E+26 | 3.9 | 1.0E+26 | 3.5 | 4.5E-02 | 1.0E-06 | |
| **InOx based interconnect** | 10 | 5.6 | 1.0E+27 | 4.0 | 1.0E+27 | 3.5 | 8.0E-04 | 8.0E-04 | |
| **Recombination interface** | | | | | | | | | |
| **PEDOT:PSS** | 10 | 5.17 | 1.0E+27 | 2.5 | 1.0E+27 | 3.5 | 8.0E-04 | 8.0E-04 | |
| **HTL/Perovskite interface** | 1 | 5.17 | 2.2E+24 | 3.9 | 2.2E+24 | 40 | 1 | 1 | |
| **perovskite** | 400 | 5.17 | 2.2E+24 | 3.9 | 2.2E+24 | 40 | 1 | 1 | 7.0E+15 |
| **ETL/Perovskite interface** | 1 | 5.17 | 2.2E+24 | 3.9 | 2.2E+24 | 40 | 1 | 1 | 1.0E+16 |
| **C60** | 30 | 5.9 | 1.0E+26 | 3.9 | 1.0E+26 | 3.5 | 4.5E-02 | 1.0E-06 | |
| **BCP** | 8 | | | | | | | | |
| **Ag** | 100 | | | | | | | | |

| Parameter | Electron capture rate | Hole capture rate | Trap energy depth | Hole trap density | Electron capture rate | Hole capture rate | Trap energy depth | work function | acceptor doping | donor doping |
|---|---|---|---|---|---|---|---|---|---|---|
| Unit | $cm^3/s$ | $cm^3/s$ | eV | $cm^{-3}$ | $cm^3/s$ | $cm^3/s$ | eV | eV | $cm^{-3}$ | $cm^{-3}$ |
| **Layers** | | | | | | | | | | |
| **ITO** | | | | | | | | 5.5 | | |
| **PTAA** | | | | | | | | | | |
| **HTL/Perovskite interface** | | | 0.80 | | 1.0E-08 | 1.0E-08 | 0.80 | | | |
| **Perovskite** | 5.0E-09 | 5.0E-09 | 0.80 | | 5.0E-09 | 5.0E-09 | 0.80 | | | |
| **ETL/Perovskite interface** | 1.0E-08 | 1.0E-08 | 0.80 | 2.0E+17 | | | | | | |
| **C60** | | | | | | | | | | |
| **InOx based interconnect** | | | | | | | | | | 1.0E+20 |
| **Recombination interface** | | | | | | | | | | |
| **PEDOT:PSS** | | | | | | | | | 1.0E+20 | |
| **HTL/Perovskite interface** | | | 0.60 | | 1.0E-08 | 1.0E-08 | 0.60 | | | |
| **Perovskite** | 5.0E-09 | 5.0E-09 | 0.60 | | 5.0E-09 | 5.0E-09 | 0.60 | | | |
| **ETL/Perovskite interface** | 1.0E-08 | 1.0E-08 | 0.60 | | | | | | | |
| **C60** | | | | | | | | | | |
| **BCP** | | | | | | | | 3.9 | | |
| **Ag** | | | | | | | | | | |

## Supplemental Note 1:

In case of the 1.80/1.27eV tandem combination, the integrated EQE of the LG subcell is significantly lower than the corresponding HG subcell leading to a current mismatched device, see also figure S18. Although there is no current matching, the performance and $J_{SC}$ of the devices - measured using an additional illumination mask to avoid parasitic effects and avoid overestimation of $J_{SC}$ - is still good. We believe this is due to the fact that the LG subcell has a low shunt resistance – which allow for the tandem to operate without strict current matching. To substantiate our hypothesis we measured sub-cell selective resistive photovoltage (RPV) of our perovskite tandem and corresponding single junction devices. In this measurement samples are excited with a 5ns long laser pulse and the photovoltage is recorded as a function of time. Using a load resistance of 1MΩ, and a correspondingly long RC time means that charge carriers will accumulate at the respective electrodes after transit through the whole device, allowing us to extract and compare transit times. As shown in **Figure S15 a and b**, we observe relatively fast and comparable transit times (~$10^{-7}$s) in HG and LG perovskite single junctions. We then measured RPV of the individual subcells, using appropriate laser wavelengths that are selectively absorbed. In this case, we expect longer transit times, as we have a much thicker layer stack the charge carriers need to transit through. Indeed, we observe longer transit times when exciting the PbSn subcell. However, the transit time remains short when exciting the HG subcell, indicating some shunts within the PbSn subcell, leading to a faster photovoltage buildup.

To corroborate this hypothesis further we performed electrical simulations in LTspice using a simplistic equivalent circuit comprising two diodes with parallel shunt resistances. A relatively low shunt resistance in the range of 0.1-5 kΩ/cm$^2$ hereby allows to accurately describe the observed *JV* characteristics and reproduces our observation - that the $J_{SC}$ must not be limited by the normally current limiting subcell in case of pinholes and/or low shunt resistances. We show simulations for various shunt resistances, in the SI, **Figure S16**, and further note that this phenomenon lowers the fill factor (FF) and thus may rather reduce the PCE compared to shunt-free current-matched devices. The effect therefore represents a loss mechanism, limiting performance, and consequently our measured PCE values are not overestimated due to this effect. Robustness against current mismatching on the other hand is highly interesting for tandem photovoltaics considering spectral variations throughout the day and hence could increase the overall energy yield in real-world applications.

In better current matched devices e.g. the 1.85/1.27 and 1.88/1.27 eV HG/LG combinations, the mismatch between integrated EQE of the HG and LG subcells is smaller, see **Figure S18**, with less impact of the above described effect.

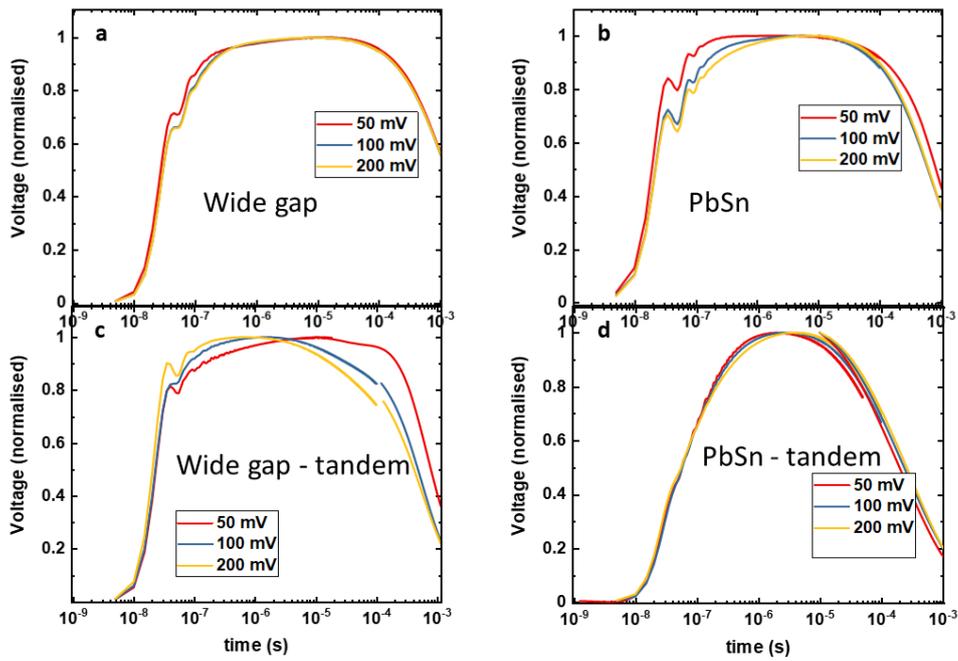

**Figure S15.** Resistance dependent Photovoltage (RPV) measurements for a) a high gap single junction, b) a low gap single junction, c) a high gap subcell in a 2T all-perovskite cell and d) a low gap subcell in a 2T all-perovskite cell.

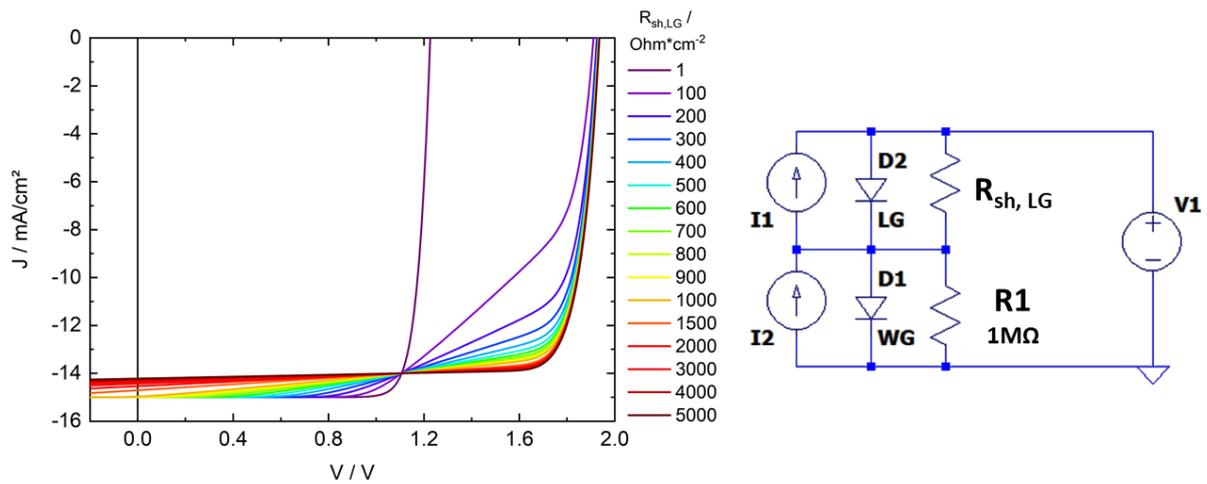

**Figure S16.** Simulations carried out using the LTspice electric circuit simulation tool to demonstrate the increasing reduction in *FF* of the tandem *JV* upon decreasing shunt resistance. On the right, the equivalent circuit used for these simulations is displayed.

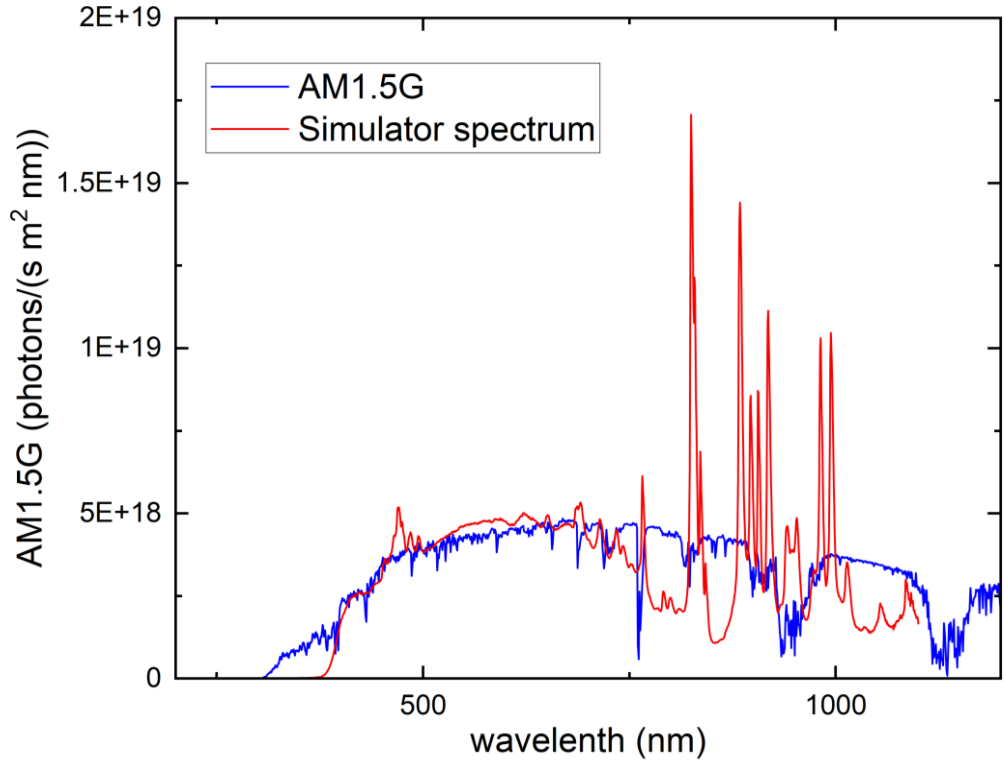

**Figure S17.** Measured solar simulator spectrum compared with the AM1.5G spectrum. Spectral mismatch values were calculated for the different HG bandgaps by integrating multiplying the measured EQE with our solar simulator spectrum and the AM1.5G spectrum, respectively, and dividing the two integrated currents obtained by each other. The spectral mismatch values obtained were 1.0019, 1.0003 and 1.0004 for HG perovskites with bandgaps of 1.80, 1.85 and 1.88 eV, respectively.

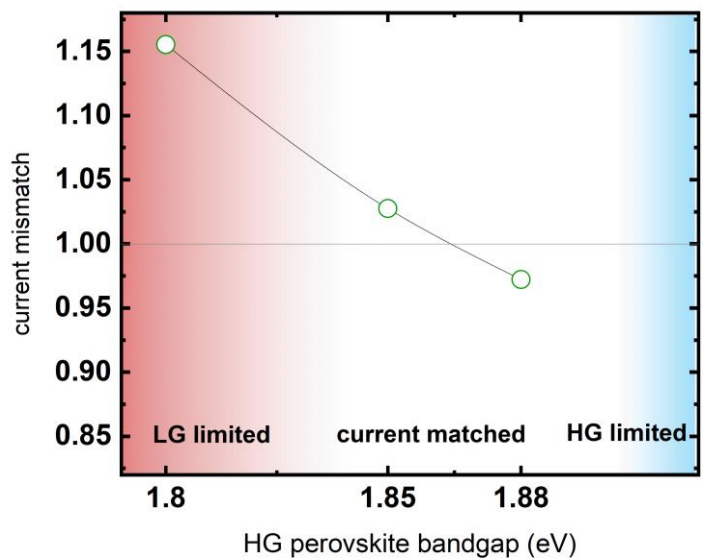

**Figure S18.** Current mismatch of optimized all-perovskite tandem solar cells based on 1.80/1.27 eV, 1.85/1.27 eV, and 1.88/1.27 eV bandgap combinations. The mismatch is calculated from the integrated EQE currents.